\documentclass{llncs}

\usepackage{amssymb}

\usepackage{amsmath}
\usepackage{mathtools}
\usepackage{algorithm}
\usepackage{algorithmic}
\usepackage{xspace}

\newcommand{\draft}[1]{}

\usepackage{url}
\usepackage[colorinlistoftodos,textwidth=60pt]{todonotes}

\usepackage{float}

\usepackage{proof}

\usepackage{algorithmic}
\algsetup{linenosize=\small}
\usepackage{setspace}

\usepackage{setspace}

\newcommand{\suppr}[1]{}

\newcommand{\scut}[1]{}

\newcommand{\powerset}{\mathcal{P}}

\newcommand{\Logicname}[1]{\ensuremath{\mathsf{#1}}}
\newcommand{\LTL}{\Logicname{LTL}\xspace}

\newcommand{\CTL}{\Logicname{CTL}\xspace}
\newcommand{\CTLs}{\Logicname{CTL^*}\xspace}
\newcommand{\ATL}{\Logicname{ATL}\xspace}
\newcommand{\ATLs}{\Logicname{ATL^*}\xspace}
\newcommand{\CTLp}{\Logicname{CTL^{+}}\xspace}
\newcommand{\ATLp}{\Logicname{ATL^{+}}\xspace}

\newcommand{\cgm}{\ensuremath{\mathcal{M}}\xspace}

\newcommand{\agents}{\ensuremath{\mathbb{A}}\xspace}
\newcommand{\coalA}{\ensuremath{A}\xspace}
\newcommand{\states}{\ensuremath{\mathbb{S}}\xspace}
\newcommand{\actions}[1]{\ensuremath{\mathsf{Act}_{#1}}\xspace}

\newcommand{\mapAct}[1]{\ensuremath{\mathsf{act}_{#1}}\xspace}

\newcommand{\plays}{\ensuremath{\mathsf{Plays}\xspace}}
\newcommand{\hists}{\ensuremath{\mathsf{Hist}\xspace}}

\newcommand{\lab}{\ensuremath{\mathsf{L}\xspace}}

\newcommand{\scomment}[1]{\textcolor{red}{\textbf {S's comment:} #1  \textbf{End of S's comment}}}

\newcommand{\ftrans}{\ensuremath{\mathsf{out}}}
\newcommand{\outcome}{\ensuremath{\mathsf{Out}}\xspace}
\newcommand{\prop}{\ensuremath{P}\xspace}
\newcommand{\flabel}{\ensuremath{L}}

\newcommand{\play}{\ensuremath{\lambda}\xspace}

\newcommand{\stratA}{F_{\hspace{-2.0pt}\coalA}}

\newcommand{\strat}[1]{F_{\hspace{-2.0pt}{#1}}}

\newcommand{\props}{\ensuremath{\mathbb{P}}\xspace}

\newcommand{\always}{\ensuremath{\Box}\xspace}
\newcommand{\event}{\ensuremath{\Diamond}\xspace}
\newcommand{\nxt}{\ensuremath{\bigcirc}\xspace}

\newcommand{\until}{\ensuremath{\mathsf{U}}\xspace}
\newcommand{\release}{\ensuremath{\mathsf{R}}\xspace}

\newcommand{\dlangle}{\langle\!\langle}
\newcommand{\drangle}{\rangle\!\rangle}
\newcommand{\diaA}{\dlangle \coalA\drangle}

\newcommand{\model}[1]{\cgm,\play_{#1}\models}
\newcommand{\mods}{\cgm,s\models}
\newcommand{\gsl}{\geq 1}

\newcommand{\fg}{\mathsf{dec}}
\newcommand{\g}[1]{\mathsf{dec}(#1)}

\newcommand{\calcul}{\mathbf{MoCOFAP }}

\newcommand{\jeter}[1]{#1}

\newcommand{\cancella}[1]{}

\begin{document}

\title{Tableaux like model checking on-the-fly for \ATLp}
\titlerunning{Tableaux like Model Checking}  
%
\author{Serenella Cerrito, serenella.cerrito@univ-evry.fr}
\authorrunning{S. Cerrito}
%
%
%
\institute{Universit\'e Paris-Saclay, Univ. Evry,  IBISC, France  }


\maketitle              

\begin{abstract}
$\;$ \\
We propose a model checking algorithm to test properties of systems that are expressed in the multi-agent temporal logic \ATLp. The specificities of this algorithm are:  it is on-the-fly, generating states only when they are needed, and it works by constructing a candidate formal proof in an inference system inspired by tableau proof systems.

\end{abstract}

\noindent
\textbf{Keywords}:   Alternating-time temporal logic,   formal verification,  model checking, multi-agent systems, proof calculi,  tableaux, temporal logics.

\section{Introduction}
\label{intro}
Alternating-time logics -- logics of the \ATL family -- are branching time temporal logics that differ from the well-known \CTL logics because they consider a multi-agent framework.
They were introduced in \cite{AHK02}.

In general,  the model checking problem amounts to computing  the truth of   a given property in an abstract structure $\cgm$ modeling a computer system, $\cgm$ being some kind of transition system. One can consider two variants of such a problem: in \textit{global model-checking} the set of states where the property holds is computed, while in  \textit{local model-checking} it is decided whether or not the property holds at a given state. Obviously the second variant corresponds to a decision problem that can be reduced  to the first formulation.

In the case of the simplest \ATL logic, named \textit{Vanilla} \ATL or simply \ATL, the global model-checking  problem has a linear time solution, both with respect to the size of the structure and the size of the formula $\varphi$ expressing the property\cite{AHK02}.

However the problem is 2EXPTIME complete for \ATLs with perfect recall strategies, that is a very expressive extension of Vanilla \ATL \cite{AHK02}, and it is PSPACE complete for the intermediate logic \ATLp (always with perfect recall strategies), that does not allow for nested temporal modalities but enables boolean combinations of temporal path formulae.  The logic \ATLp is less expressive than \ATLs but strictly more expressive than Vanilla \ATL and allows for the expression of natural properties that Vanilla \ATL cannot formulate (see for instance \cite{Bulling2010a}). 
The PSPACE complexity for \ATLp has been first stated in   \cite{Bulling2010a}; however, as observed in \cite{GS}, the proof of the upper bound  given in that work contains a flaw. A new proof of the result has been given in \cite{GS} by using a game-theoretical semantics, proved to be  equivalent to the original compositional semantics first introduced in \cite{AHK02}.
In \cite{GS} also some tractable fragments $\ATL^k$ of \ATLp, where $k$ is a positive integer, are introduced, and proved to have a PTIME complexity.

allow us to
 Already in the computationally simple case of Vanilla \ATL the size of the model actually constitutes a problem in practical cases, because the global number of states can be really big (state explosion problem). \textit{In this work we propose an \underline{on-the-fly} local model checking algorithm for \ATLp with perfect recall strategies, that is founded on $\calcul$, a tableau-like inference system that we define here}. 
 
 The general practical advantages of on-the-fly model checking algorithms to decide the local model-checking problem are well-known:  they allow us to work with the intensional, implicit description of the model  and for  the explicit construction of states of the model  only when they are needed to evaluate the input formula $\varphi$ at given state $s$. Hence evaluation is oriented by an analysis of $\varphi$ and, in the general case, only a subset of the states that are reachable from $s$ is actually built. Up to our knowledge this is the first on-the-fly model checking algorithm for  logics of the \ATL family.


The outline of this work is the following. In Section \ref{pre} we recall the syntax and the (compositional) semantics of \ATLs and \ATLp.
Our approach to local model checking for \ATLp is developed in Section \ref{approach} where the calculus $\calcul$ is described and its main properties are
formulated.\footnote{Complete proofs of the results can be found in the Appendix, that will be removed in case of acceptance of the present paper to FORMATS.}
Section \ref{conclu} concludes this work, by discussing its connection with some related literature and by pointing at some issues that still need to be explored and that will be the subject of future research.

\section{Preliminaries}
\label{pre}
We recall here some standard definitions about  logics of the \ATL family (see for instance \cite{livreATL} for more details).
\begin{definition}[Concurrent Game Model] 
\label{cgm}
Given  a set of \emph{atomic propositions} $P$, a CGM
 (\textit{Concurrent Game Model}) is a  5-tuple 
$$\cgm = \langle   \agents,  \states, \{\actions{a}\}_{a \in \agents}, \{\mapAct{a}\}_{a \in \agents}, \ftrans, \lab \rangle$$ such that:\\
$\bullet$ \agents = $\{1, ..., k\}$  is a finite non-empty set of \textit{agents};\\
$\bullet$ \states is a non-empty set of \textit{states};\\
$\bullet$ For each $a \in \agents$, \actions{a} is a non-empty set of \textit{actions}. If $A \subseteq \agents$, then $A$ is a  \textit{coalition} of agents.
Given a coalition $A$, an  $A$-move is  a  $k$-ple 
$\langle \alpha_1,...,\alpha_k \rangle$ where, for any  $i, 1 \leq i \leq k$,  if  $i  \in A$ then  $\alpha_i  \in \actions{a}$, else $\alpha_i =  *$  ($*$ being a place-holder symbol distinct from each action). A move of the coalition of all the agents, $\agents $, will also be called \textit{global move}. The  set  of  all the  $A$-moves is denoted by  $\actions{A}$.  The  notation  $\sigma_A$  denotes an element of $\actions{A}$,  and  if $a \in A$, $\sigma_A(a)$ means the action of the agent  $a$ in the $A$-move $\sigma_A$;\\
$\bullet $ \mapAct{a} is  a function  mapping  a  state  $s$ to  a non-empty subset of  \actions{a}; $\mapAct{a}(s)$ denotes  
the set of  actions  that are available  at state $s$ to the  agent  $a$.  Given a coalition $A$, a mapping $\mapAct{A}$ associating to a state a set of $A$-moves  is  naturally induced by the function $\mapAct{a}$;  $\mapAct{A}(s)$  is the set of all the $A$-moves available to coalition  $A$  at state $s$. \\
$\bullet$  $\ftrans \; $ is a \textit{transition function}, associating to each   $s \in \states$ 
and each 
$\sigma_\agents
 \in \mapAct{\agents}(s)$ a state 
$\ftrans(s, \sigma_\agents) \in \states$: the state reached when each  $a \in \agents$ does the  action $\sigma_a$
at $s$;\\
$\bullet$ 
$ \lab $ is a labelling function $\lab:\states\rightarrow \powerset(\prop)$, associating to each state $s$ the set of propositions holding at $s$.

\end{definition}

It is worthwhile observing that the above definition does not require  the set \states to be finite. In this work, however, it will  always be finite.\footnote{Observe that  the existence  of sound, complete and terminating tableaux for satisfiability testing of  \ATL*  formulae, as in \cite{davidCADE}, is also a proof of the finite model property for \ATLs, the most expressive logic of the \ATL family considered here.}

\subsection{\ATL Logics Syntax}
We consider here three logics of the $\ATL$ family that are, in order of decreasing expressivity : $\ATLs$, $\ATLp$ and $\ATL$.
An \textit{atom} $At$ is either a symbol $p \in P$ or $\top$ (True).  A \textit{literal} $l$ is either an atom or an expression $\neg At$ where $At$ is an atom. Below, $At$ is an atom and 
$A$ is a \textit{coalition} of agents. The following grammar defines formulae by mutual recursion between two sorts of them.
\begin{definition}[\ATLs syntax]
\label{atlStar-gram}
$\;$\\
$\ATLs$-state formula 
$\psi :=  At\;\mid \; (\neg \psi) \; \mid \; (\psi \wedge \psi) \;\mid \; (\psi \vee \psi)\;\mid \;(\diaA \Phi) 
 \;\mid \; ([[A]] \Phi) $\\
$\ATLs$-path formula
$
\Phi :=  \psi \;\mid \;    (\neg \Phi )\;\mid \; (\Phi \wedge \Phi) \; \mid \; (\Phi \vee \Phi) \;\mid \;(\nxt \Phi) \;\mid \;(\always \Phi) \;\mid \;(\Phi \until \Phi)
$
\end{definition}

The expressions $\diaA$ and $[[A]]$ are called   \textit{strategic quantifiers} (respectively: existential and universal), while $\nxt, \always$ and $\until$ are \textit{temporal operators}
(already used in \LTL, \textit{Linear Temporal Logic}, see \cite{livreATL} for instance).
It is worthwhile observing that  any   \ATLs state formula is also an   \ATLs  path formula, while the converse is false. State formulae will always be noted by lower case Greek letters, and path formulae by upper case  Greek letters. 

When the syntax of formulae is restricted so as to impose that any strategic temporal operator is always immediately dominated by a strategic quantifier $Q$, and a strategic quantifier $Q$ always immediately dominates a temporal operator, so as to obtain syntactic undecomposable blocks as $Q \; \nxt$, $Q \; \always$ and $Q \; ... \until \; ...$, then one obtains the strictly less expressive  but computationally easier logic \textit{Vanilla \ATL}, often called just \ATL.  
\textit{The logic $\ATLp$ is intermediate between $\ATLs$ and $\ATL$}. \footnote{A similar relation holds between the three computation tree temporal logics $\CTLs$, $\CTL$ and $\CTLp$ : see \cite{livreATL}} Its grammar differs from the $\ATLs$ one just \underline{with respect to path formulae}.  An \ATLp path formula $\Phi$ is either a state formula or the negation of
a $\Phi$ formula or a conjunction of $\Phi$ formulae or applies temporal operators to  \underline{state} formulae.
\jeter{\begin{definition}[\ATLp path formulae]
\label{atlPlus-gram}
$\;$\\
$\ATLp$-path formula
$
\Phi :=  \psi \;\mid \;  (\neg \Phi )\;\mid \; (\Phi \wedge \Phi) \;\mid \;(\nxt \phi) \;\mid \;(\always \phi) \;\mid \;(\phi \until \phi)
$
\end{definition}
}

The path formula $\event \Phi$ can be  defined by $T \until \Phi$.
In the case of $\ATLs$ and $\ATLp$   the  universal strategic quantifier can be expressed   by means of the existential one: $[[A]] \Phi \; \equiv \; \neg [[A]]  \neg \Phi$; moreover the $\LTL$ temporal operator $\release$ (\textit{release}) can be defined via $\always$ and $\until$: $\Phi \release \Psi$ $  \equiv \; (\always \Psi) \vee (\Phi \until (\Phi \wedge \Psi))$. Because of the well-known  $\LTL$ equivalence  $\neg (\Phi \until \Psi) \; \equiv \; (\neg \Phi) \release (\neg \Psi)$, it follows that 
$\ATLs$ and $\ATLp$ formulae can always be rewritten in \textit{negation normal form} \textit{(fnn)}, where negation applies only to atoms. \footnote{In \cite{Lar2007} it is stated that Vanilla  \ATL doest not enjoy negation normal form. One can observe, however, that this wouldn't be true if the \textit{release} operator were given as a primitive modality in the grammar.}

In the sequel we always  consider \ATLs (and \ATLp) formulae that are in \underline{negation normal form}, since this is practical for the definition of the calculus $\calcul$ in Section \ref{approach}. 


\subsubsection{Classification of \ATLs state formulae.}
 Let us denote by $Q$ a strategy quantifier. Following \cite{ijcar14,atlplusjournal,davidCADE} 
\ATLs   formulae can be partitioned as follows. 

\begin{itemize}
\item A \textit{successor formula} is a formula having either  the form 
$\diaA \nxt  \Phi$ or 
$ [[A]] \nxt \Phi$. 
\item Then:
\begin{itemize}

\item 
\textit{Primitive Formulae}. A primitive formula is either a literal or a successor formula $Q \nxt  \phi$ where  $\phi$ is a \underline{state} formula, called the successor component of $Q \nxt  \phi$\footnote{Observe that in the case of \ATLp any successor formula  is necessarily  primitive.
};\\
\item
  \textit{$\alpha$-formulae} are state formulae of the form $\phi \wedge \psi$; $\phi$ and $\psi$ are said to be its $\alpha$-components;\\
\item 
\textit{$\beta$-formulae}, are state formulae of  the form $\phi \vee \psi$; $\phi$ and $\psi$ are said to be its $\beta$-components;\\
\item
  \textit{$\gamma$-formulae} are  state non-primitive formulae  $Q\;    \Phi$.\\ 
\end{itemize}
\end{itemize}

\begin{remark}
\label{menomale}
Given the structure of $\ATLp$ path-formulae, any $\gamma$-formula for $\ATLp$ has the form $Q\;    \Phi$ where  $Q$ is a strategic quantifier and $\Phi$ is a boolean combination of sub-formulae having 
one of the following forms $\psi_1$, $\nxt \psi_1, \always \psi_1, \nxt \psi_1, \psi_1 \until \psi_2$, where $\psi_1, \psi_2$ are \underline{state} formulae. 

\end{remark}


\subsection{Semantics}
The semantics for \ATLs is based on the notions of concurrent game model, \textit{play} and \textit{strategy}.

A play    $\lambda$ in a CGM \cgm is an infinite sequence of elements of $\states$: $s_0, s_1, s_2,...$ such that for every 
 $i \geq 0$,  there is  a global move $\sigma_{\agents} \in \mapAct{\agents}(s_i)$ such that $\ftrans(s_i, \sigma_{\agents})= s_{i+1}$. Given a play $\lambda$, we denote by $\lambda_0$ its initial state, by $\lambda_i$ its $(i+1)$th state, by $\lambda_{\leq i}$ the prefix $\lambda_0 ... \lambda_i$ of $\lambda$ and by $\lambda_{\ge i}$ the suffix $\lambda_i \lambda_{i+1} ...$ of $\lambda$. 
 \jeter{Given a prefix $\lambda_{\leq i}:\lambda_0 ... \lambda_{i}$, we say that it has length $i+1$ and write $|\lambda_{\leq i}| = i+1$.  An empty  prefix has length 0.}
 A (non-empty) \emph{history} at state $s$ is a finite prefix of a play ending with $s$.  
We denote by $\plays_\cgm$ and $\hists_\cgm$ respectively the set of plays and set of histories in a CGM $\cgm$. 
\jeter{For a state $s \in \states$ we note $\plays_\cgm(s)$ and $\hists_\cgm(s)$ as the set of plays and set of histories with initial state $s$.} 

Given a coalition $A \subseteq \agents$ of agents,  a \textit{perfect recall 
$A$-strategy} $\stratA$ is a function which maps each element $\lambda = \lambda_0 ... \lambda_\ell $ of $\hists_\cgm$  
to an $A$-move $\sigma_{A}$ belonging  to
 $\mapAct{A}(\lambda_\ell)$ (the set of actions available to $A$  at state $\lambda_\ell$). Whenever  $\stratA$ depends only  on the state  $\lambda_\ell$ the strategy is said to be  \textit{positional}.  The \textit{complementary coalition of $A$} is the coalition $\agents \setminus A$. A \textit{perfect recall $A$-co-strategy} is a function which maps each element $\lambda = \lambda_0 ... \lambda_\ell $ of $\hists_\cgm$  
to a $C$-move $\sigma_{C}$  of $C$, the complementary coalition of $A$, belonging  to
 $\mapAct{C}(\lambda_\ell)$ (the set of actions available to $C$  at state $\lambda_\ell$).  We will note ${\stratA}^c$ a $A$-co-strategy.
 \textit{In  the rest of the paper we always consider perfect recall strategies (and co-strategies)}.

For any coalition $A$, a global move $\sigma_\agents$ \textit{extends} an $A$-move $\sigma_A$ whenever    for each agent 
$a \in A$,  $\sigma_A(a) = \sigma_\agents(a)$. 
Let $\sigma_A$ be an $A$-move; the notation 
$\outcome(s, \sigma_A)$ denotes the set of states 
$\ftrans(s, \sigma_\agents)$ where $\sigma_\agents$ is any global  move
extending $\sigma_A$. Intuitively, $\outcome(s, \sigma_A)$  denotes the set of the states that are successors of $s$ when the coalitions $A$ plays at $s$  the $A$-move  $\sigma_A$ and the other agents play no matter which move.

A play  $\lambda = \lambda_0, \lambda_{1},...$ is said to be \textit{compliant with a strategy $\stratA$} if and only if  for each 
$j \geq 0$ we have $\lambda_{j+1} \in \outcome( \lambda_j, \sigma_A)$, where $\sigma_A$ is the  $A$-move
chosen by  $\stratA$ at state $\lambda_i$.

Below we recall the compositional semantics introduced in \cite{AHK02}.
 
The notion \textit{\cgm satisfies the formula $\Phi$ at state $s$}, noted $\mods\Phi$, is defined by induction on $\phi$ as follows (omitting the obvious boolean cases):
\begin{itemize}
\itemsep 0 cm 
\item $\mods \top$ holds always and  $\mods  \neg \top$ never holds;
\item $\mods p$ iff $p \in \flabel(s)$, for any proposition $p \in \props$;
\item $\mods \neg p$ iff $p \not \in \flabel(s)$, for any proposition $p \in \props$;
\item $\mods\diaA\Phi$ iff there exists an \coalA-strategy $\stratA$ such that, for all plays $\play$ starting at $s$ and compliant with the strategy $\strat{A}$, $\model{}\Phi$;
\item $\mods[[A]]\Phi$ iff there exists a co-$A$-strategy  ${\stratA}^c$ such that, for all plays $\play$ starting at $s$ and compliant with the strategy ${\strat{A}}^c$, $\model{}\Phi$;
\item $\model{}\varphi$ iff $\model{0}\varphi$;
\item $\model{}\nxt\Phi$ iff $\model{\gsl} \Phi$;
\item $\model{}\always\Phi$ iff $\model{\geq i} \Phi$ for all $i \geq 0$;
\item $\model{}\Phi\until\Psi$ iff there exists an $i\geq 0$ such that:  $\model{\geq i}\Psi$ and for all $0 \leq j < i$, $\model{\geq j}\Phi$.
\end{itemize}
Given a CGM \cgm, a state $s$ of $\cgm$ and a state formula $\varphi$,  we say   that \textit{$\varphi$ is true at $s$} whenever  $\cgm,s \models \varphi$.

\begin{example}
\label{exa1}
Figure \ref{charriots-robot} shows a simple but classical example of CGM (with just two agents), borrowed from \cite{Bulling2010a}. We have two agents that are two robots, say 1 and 2,  that can push a charriot in two opposite directions, or do nothing (wait). The charriot can be in three possible positions, corresponding to three states $q_0, q_1$ and $q_2$ of the CGM that are characterized  by which one of three boolean variables, $p_0$, $p_1$ and $p_2$ is true. Observe that at $q_0$ the formula $\langle  1 \rangle \nxt pos_2$ is false while  both the formulae $\langle  1 \rangle \nxt  (pos_0 \vee ps_1 \vee pos_2)$  and  $\langle  1,2 \rangle \nxt pos_2$ are true.

\begin{figure}[h]
\label{charriots-robot}
\includegraphics[scale=0.32]{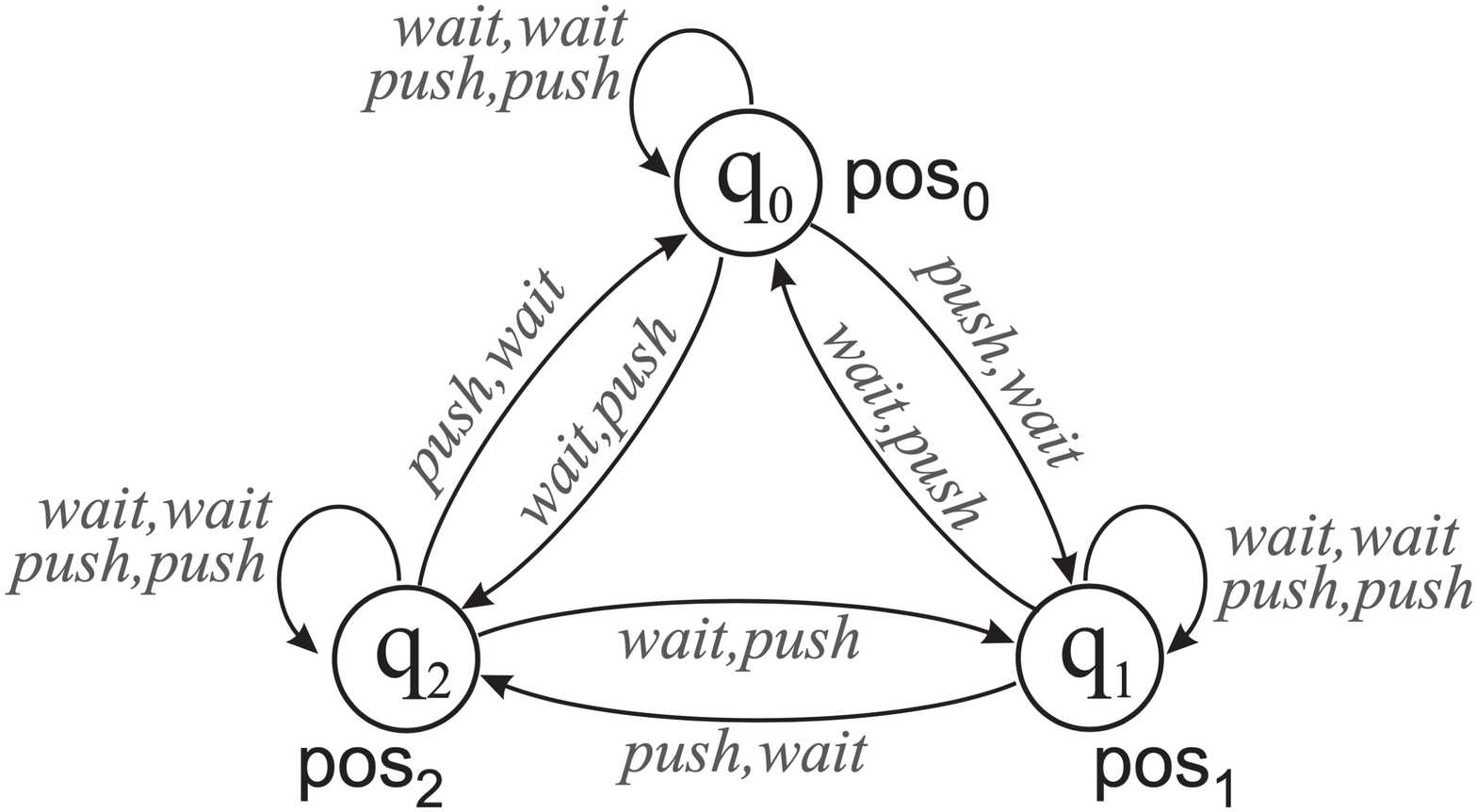}
\caption{Two robots and a carriage : a CGM modeling the scenario}

\end{figure}
\end{example}

 Given any pair of  path formulae $\Phi$ and $\Psi$, $\Phi$ is a \textit{logical consequence} of $\Psi$, noted $\Psi \models \Phi$, when for any CGM $\cgm$ and any path $\lambda$, $\model{} \Psi$
  implies   $\model{} \Phi$; the formulae $\Psi$ and $\Phi$ are said to be \textit{equivalent}, noted $\Psi \; \equiv \; \Phi$ when $\Psi \models \Phi$
and $\Phi \models \Psi$.
\jeter{Sometimes  Definition \ref{cgm} is completed by specifying a set of \textit{initial states} and model-checking a formula $\phi$ on $\cgm$  means to test whether
$\phi$ is true at each initial state.}
The following fixed-point equivalences hold for \ATLs with perfect recall strategies (\cite{livreATL}):
\begin{itemize}
\item $\diaA \always \phi  \; \equiv \; \phi \wedge \diaA \nxt \diaA \always \phi $.
\item $\diaA \phi  \until \psi \; \equiv \; \psi \vee (\phi \wedge \diaA \nxt \diaA
\phi  \until \psi)$ 
\item The analogous of the equivalences above with 
$[[A]]$ replacing  $\diaA$.
\end{itemize}

\section{Our Approach to Model Checking  \ATLp properties on-the-fly}
\label{approach}
\subsection{Global Overview of Our Algorithm}
\textit{In this work we  propose an algorithm to decide on-the-fly whether a given \ATLp state formula $\varphi$ is true at given state $s$ of a given CGM \cgm  or not}.  To our knowledge this is the first on-the-fly algorithm for \ATLp. It works  by inspecting whether a candidate proof  of such a truth   -- in 
the formal system $\calcul$ that we propose --
is successful or not, and is indeed a proof. Given $\cgm$, a candidate proof for $\varphi$ and  $s$ here  is a possibly infinite tree (encoperformanceded by a finite directed graph with cycles) where each node is labeled by a set of assertions $S$  interpreted \underline{disjunctively},
that is, true whenever at least one of its elements holds. Each assertion will say that   a given  \ATLp (state)  formula $\psi$ holds at a given state, and the root will express that $\cgm, s \models \varphi$.
Roughly, a candidate proof $\cal P$ is indeed a proof of its root if and only if \underline{each} branch is successful, i.e. either is finite and ends with $T$ ($True$), or else is infinite but it contains a ``virtuous" circle
(see later for precise definitions). One can finitely detect whether a branch is successful or not, and for infinite branches this can be efficiently done by applying Tarjan's algorithm to find strongly connected components in a directed graph \cite{Tarjan}.  A node of the tree may have $k$ children, $k \geq 0$, generated by applying to the parent node  an inference 
rule of the calculus $\calcul$ defined below. 
Each rule (but $(True)$, see later on) is such that its premise is true if and only if {all} the  conclusions are true, that is, branching in the tree  is interpreted \underline{conjunctively}. \textit{This is essential to our approach.} In fact, our decision algorithm  builds \underline{depth-first} a candidate proof  tree  and interrupts the construction of such a tree as soon as a branch is recognized to be unsuccessful:  the root is then declared to be false. Of course, whenever $\varphi$ holds at $s$, all the branches of the tree need to be constructed in order to conclude to the truth of the root. But in practical cases early halting the construction of a candidate proof of a false assertion  is quite useful and can significantly reduce the space of the  states of $\cgm$ that need to be built.\footnote{If the problem were solved by reducing it to global model-checking,  whenever $\cgm, s \not \models \varphi$ all the states of $\cgm$ would be explored, to finally find out 
that $s$ is not an element of the set of states satisfying $\varphi$.}

Our approach is inspired by tableaux methods, generally used, however,  to prove or refute the satisfiability of formulae, not to do model-checking. The extension of our approach to $\ATLs$ is ongoing work. In the next section we describe our  calculus, named  $\calcul$ as   \textit{\textbf{Mo}del \textbf{C}hecking \textbf{O}n-the-\textbf{F}ly for \ATLp}.

\subsection{Rules of the calculus $\calcul$}
\label{sec_calcul}

\medskip

\noindent
 \textbf{Assertions.} 
Given a CGM, we will use nominals $s_1, s_2,...$ to name its states. In the sequel by an abuse of language we identify nominals and states that they denote.
\begin{definition}
\label{def_assertion}
Let $\cgm$ be a CGM and $s$ one of its states.
An assertion is an expression  of the form  $s \vdash_{\cgm} \phi$ 
where $\phi$ is a state \ATLp formula.\\
An assertion   $s \vdash_{\cgm} \phi$ is true whenever
$  \mods \phi$ holds.

\end{definition}

Let us emphasize that assertions provide a  syntax that internalizes in our calculus the  meta-theoretic notion of truth of an
\ATLp formula at a state, as defined in  Section \ref{pre}. Our calculus will allow one to syntactically prove (or refute) an assertion saying  that a given  formula is true at a given state.
We will use lower case roman letters $a, b, c...$ to denote assertions and
upper case roman letters $A, B, C,...$ to denote sets of assertions. 
\begin{definition}
\label{assertion_set}
A finite set $A= \{a_1,...,a_n\}$ of assertions   (a clause) referring to  a given CGM
is true when at least  one $a_i$ is true. 
\end{definition}
As a consequence of the above definition when we write $\{a_1,...,a_n\}$
the comma has the semantics of a \underline{disjunction} over assertions (not to be confused with the disjunction operator  $\vee$ of formulae), and the empty set of assertions is always false. The notation $T$ will be used to denote a set of assertions which is always true.

\noindent
\textbf{Expansion rules.}


\noindent
The rules of $\calcul$ allowing one to prove (or refute) that a given assertion
(or a given set of assertions)  is true   will be presented in the form:
$$ \frac{A}{B_1  \; \;   \& \; ... \; \& \; \;  B_n}$$
where $A$ and the $B_i$, $1 \leq  n$  are sets of assertions. 
The set $A$ is the \textit{premise} of the rule and $B_1$,..$B_n$ are called \textit{conclusions} or
\textit{expansions}.
Differently from usual tableau calculi aiming at testing satisfiability, as for instance the tableaux in \cite{atlplusjournal,davidCADE} for \ATLp formulae, where the satisfiability of the premise implies the satisfiability of at least one expansion,  rules of $\calcul$ are designed so that 
if  the premise is true then  \underline{all} the expansions  are true.
In those works  rules are such that  expansions are disjunctively connected (and comma in any expansion has a conjunctive meaning), while in our calculus  expansions are \underline{conjunctively} connected (and comma in any expansion has a disjunctive meaning).

Rules will be classified as \textit{static rules} or \textit{dynamic  rule}.

\medskip

\noindent
\underline{\textit{Static Rules.}}
Each static rule has exactly one \textit{principal assertion}, that is
explicitly shown in the premise, and all the other assertions in the premise are said to constitute the  \textit{context}. 
The role of a static rule is to analyse what it means for the principal assertion to be true.

 The cases where  the principal assertion has the form 
$s \vdash_\cgm \varphi$ where $\varphi$ is a literal or an $\alpha$-formula or a $\beta$-formula are easy, and the corresponding rules are shown in Figure \ref{easy_rules}. In that figure, $E$ is a set of assertions,  $l$ is a literal, $\varphi_1$ and $\varphi_2$ are \ATLp state formulae. As usual for tableau rules, there and in the sequel of the paper we do not explicity write brackets for sets and
$E, a$, where $a$ is an assertion, means the union of $E$ and $\{a\}$.
\begin{figure}[h]
\vspace{-3mm}
\caption{Rules for literals, $\alpha$-formulae, and  $\beta$-formulae}
\label{easy_rules}
\[
\begin{array}{|l|c|l|}
\hline
&  & \\
 \infer[(True)]{T}{s \vdash_{\cgm} l, \; E}  \;\;\; if \;  \cgm, s \models l  & \;\;\;\;\;\;\; &   \infer[(False)]{E}{s \vdash_{\cgm}  l, \;  E} \;\;\; if \;  \cgm, s \not \models l  \\
\hline 
& & \\
\infer[(\alpha)]{s \vdash_{\cgm} \phi_1,   E \;\;\;\ \& \; \;  \;\   s \vdash_{\cgm}  \phi_2, E }{s \vdash_{\cgm} \phi_1 \wedge \phi_2, literal\; E } & & \infer[(\beta)]{s \vdash_{\cgm} \phi_1,  s \vdash_{\cgm}  \phi_2, \; E }{ s \vdash_{\cgm} \phi_1 \vee \phi_2, \; E } \\
\hline 
\end{array}
\]
\end{figure}
The analysis of $\gamma$-formulae is more delicate. Let us observe that  any strategic quantifier actually combines an existential and a universal quantification. For instance, $\diaA \Phi$ means that \textit{there is a strategy} for  coalition $A$  such that, \textit{no matter how the other agents play},  $\Phi$ is assured. Thus, two difficulties have to be faced, in the general case of \ATLs formulae:
\begin{enumerate}
\item 
No strategic quantifier 
distributes over conjunction, and no strategic quantifier  distributes over disjunction; 
\item 

When $\Phi$ and $\Psi$ are not state formulae, fixed point equivalences cannot be immediately exploited to analyse assertions of the form $s \vdash_{\cgm} Q \always \Phi$ and $Q \Phi \until \Psi$, where $Q$ is a strategic quantifier. Take, for instance, the ($\gamma$) \ATLs formula $\diaA \always \event p$. Its truth at a state $s$ cannot be analyzed as the truth at $s$ of both $\event p$ and $\diaA \nxt \diaA \always \event p$, because $\event p$ is not a state formula.
\end{enumerate}
In this paper, however, we deal only with \ATLp model-checking, thus we need to face only the first of the two difficulties.
In \cite{ijcar14,atlplusjournal} an approach is proposed to deal with it in the context of tableau calculi aiming at deciding the satisfiability of  \ATLp  formulae
(later extended to \ATLs  in \cite{davidCADE}). There,
any 
$\gamma$-formula is  analyzed as a disjunction of conjunctions of state formulae by means of an \textit{iteration} of \ATLs equivalences.  For  instance,
the  truth of the \ATLp $\gamma$-formula $\diaA ((\always p) \vee (\always q))$   at a current state $s$ is analyzed in three cases:\\
1)  At present state $s$,  $p$ holds (\textit{present} state formula),  and
the state formula  $\diaA \nxt \diaA (\always p)$ holds (\textit{future} state formula); this last express a commitment for the future: for some $A$-strategy the  paths issued from $s$'s successors   and compliant with the strategy will make 
the path formula $ \always p$ true;\\
2) At present state $s$,  $q$ holds (\textit{present} state formula),  and
the state formula $\diaA \nxt \diaA \always q$ holds (\textit{future} state formula), expressing the commitment that for some $A$-strategy the  paths issued from $s$'s successors   and compliant with the strategy will make 
the path formula $ \always q$ true ;\\
3) At present state $s$,  both $p$ and $q$ hold (\textit{present} state formulae),  and the state formula 
$\diaA \nxt \diaA  ((\always p) \vee (\always q))$ holds (\textit{future} state formula), expressing the commitment that for some $A$-strategy the  paths issued from $s$'s successors   and compliant with the strategy will make 
the path formula $ (\always p) \vee  (\always q)$ true. In this last case at the present state no choice has been made yet
about which one, among   $\always p$ and $\always q$, to enforce.

Hence, expansions of a vertex containing the formula $\diaA (\always p ) \vee (\always q))$ are built by exploiting its  equivalence with: \\
  $(p \wedge \diaA \nxt \diaA \always p) \vee  (q \wedge \diaA \nxt \diaA \always q) \vee (p \wedge q \wedge \diaA \nxt \diaA \always( p \vee q))$.

In the general case, the approach in \cite{ijcar14,atlplusjournal} analyses $\gamma$-formulae in \ATLp by using a function $\fg$ which takes as input a path formula  and returns a set of ordered pairs of formulae, the first being a state formula and the second a path formula. The definition of $\fg$ uses two  
auxiliary binary functions $\otimes$ and $\oplus$ taking as arguments
two sets of ordered pairs of formulae and producing a new set of ordered pairs. The first element of each pair corresponds to the \textit{present} formula, the second to the commitment for the \textit{future}.  Our approach exploits this idea by adapting it to our case, where we need the right part of equivalences as the one exemplified above to be in conjunctive normal form rather than in disjunctive normal form (with respect to primitive formulae).

\begin{definition}
\label{deco_amelie}
$\;$

\noindent
\begin{enumerate}
\item
 Let $\Gamma_1$ and $\Gamma_2$ be two sets of ordered pairs of $\ATLp$  formulae. Then:\\
\begin{itemize}
\item

$\Gamma_1 \otimes \Gamma_2$ :=
$\{ \langle \psi_i {\wedge} \psi_ j, \Psi_i {\wedge} \Psi_j \rangle  \; \mid \; \langle \psi_i, \Psi_i \rangle \in \Gamma_1,  \langle \psi_j, \Psi_j \rangle \in \Gamma_2
\}$. \\
\item
$\Gamma_1 \oplus \Gamma_2 :=$ \\
$\{ \langle \psi_i  {\wedge} \psi_ j, \Psi_i {\vee}\Psi_j \rangle  \; \mid \; \langle \psi_i, \Psi_i \rangle \in \Gamma_1, \;  \langle \psi_j, \Psi_j \rangle \in \Gamma_2, \; \Psi_i \not = \top,  \Psi_j \not = \top. 
\}$\\

\end{itemize}
{where the operators $\otimes$ and $\oplus$ are associative, up to logical equivalence.\\}

\item
\begin{itemize}
\item
The function $\fg$ 
is defined by recursion on $\Phi$:\\
\begin{itemize}
\item 

$\g{\varphi} = \{\langle \varphi, \top \rangle\}$ and $\fg(\nxt \varphi) = \{\langle \top, \varphi \rangle\}$,  for any $\ATLp$ state formula.\\
\item

$\g{\Box \varphi_1} = \{\langle \varphi_1,\Box\varphi_1 \rangle \}$ \\
\item 

$\g{\varphi_1 \until \varphi_2} = \{\langle \varphi_1 ,  \varphi_1 \until \varphi_2 \rangle,  \; \langle \varphi_2, \top \rangle\}$\\
\item
 $\g{\Phi_1 \wedge \Phi_2} = \fg(\Phi_1) \otimes \fg(\Phi_2)$\\
\item  

$\g{\Phi_1 \vee \Phi_2} = \fg(\Phi_1) \cup \fg(\Phi_2) \cup (\fg(\Phi_1) \oplus \fg(\Phi_2))$.\\ 
\end{itemize} 

\end{itemize}
\item
Let $\theta = \diaA \Phi$ or $[[A]] \Phi$ be a $\gamma$-formula. All ordered pairs $\langle \psi, \Psi \rangle$ in $\fg(\Phi)$ are converted to a state formula $\gamma_c(\psi, \Psi)$, called a $\gamma$ component of $\Theta$ as follows:\\
\begin{itemize}
\item 
$\gamma_c(\psi, \Psi) = \psi$ if $\Psi$ is $\top$\\
\item
$\gamma_c(\psi, \Psi) = \psi \wedge \diaA \nxt \diaA  \Psi$ if $\theta = \diaA \Phi$ \\
\item
 $\gamma_c(\psi, \Psi) = \psi \wedge [[A]] \nxt [[A]] \Psi$ if $\theta = [[A]] \nxt [[A]] \Phi$ if $\theta = [[A]]   \Phi$ 

\end{itemize}

\end{enumerate}

\end{definition}


\begin{definition}

\label{deco_serena}
The (conjunctive) analysis of a  state $\gamma$-formula $\varphi$ in \ATLp, noted $An(\varphi)$, is the\footnote{Unique modulo logical equivalence: for instance 
 the conjunctive normal form of $(p \vee (q \wedge \top))$ is $(p \vee q) \wedge (p \vee \top)$=$(p \vee q) \wedge \top$ = $p \vee q$.}  conjunctive normal form of the disjunction of all its $\gamma$ components.\\
 For any $\gamma$-formula $Q \Phi$, let $An(Q \Phi)$ be   $\delta_1 \wedge...\wedge \delta_n$ where each $\delta_i$ has the form ${\varphi_i}^1 \vee ... \vee {\varphi_i}^p$; we 
say that each $\delta_i$ is a 
 \textit{$\gamma_d$ component}  of $Q \Phi$.

\end{definition}
\jeter{
As an example, consider again the  
$\gamma$--formula $\diaA (\always p) \vee (\always q)$.
Here, $\fg((\always p) \vee (\always q))$=$\fg(\always p) \cup \fg(\always q) \cup (\fg(\always p) \oplus \fg(\always q))$ = $\{\langle p, \always p\rangle, 
\langle q, \always q\rangle, \langle p\wedge q, (\always p) \vee  (\always q) \rangle\}$. 
We get three $\gamma$-components:\\
 $p \wedge \diaA \nxt \diaA \always p$, \\
 $q \wedge \diaA \nxt \diaA \always q$, \\
 and $p \wedge q \wedge \diaA \nxt \diaA  (\always p) \vee (\always q)$.\\ Here $An(\diaA (\always p) \vee (\always q))$ is the conjunctive normal form of the disjunction of the three 
 $\gamma$-components.
 }
 \jeter{
A lemma in  \cite{atlplusjournal} (Lemma 3.1) states that each formula $ \diaA \Phi$ in \ATLp  is logically equivalent to the disjunction of its $\gamma$ components, and analogously for $ [[A]] \Phi$. As 
a consequence :
\begin{lemma}
\label{myDecoLemma}

Any $\ATLp$ state formula $\varphi$ is logically equivalent to $An(\varphi)$, which is a conjunction of disjunctions of primitive formulae.
\end{lemma}
}

  The $(\gamma)$-rule is formulated in Figure \ref{gamma_rule}. 
 \jeter{Because of Lemma 3.1 in  \cite{atlplusjournal}
 we can conclude:
 
 \begin{lemma}
 \label{lemma_amelie_version_S}
 The premise of a $\gamma$-rule is logically equivalent to the conjunction of its conclusions.
 \end{lemma}
 
 }
  \begin{figure}[h]
\vspace{-3mm}
\label{gamma_rule}
\caption{The $\gamma$-rule}
\label{gamma_rule}
\[
\infer[(\gamma)]{
s\vdash_\cgm {\varphi_1}^1, ..., s\vdash_\cgm {\varphi_1}^{k(1)}, E||...|| s\vdash_\cgm {\varphi_n}^1, ..., s\vdash_\cgm {\varphi_n}^{k(n)}, E
}
{s\vdash_\cgm Q \Phi, E}
\]

\medskip

\noindent
where for $1 \leq j \leq n$,  ${\varphi_j}^1 \vee ... \vee  {\varphi_j}^{k(j)}$ is $\delta_i$, as given in Definition \ref{deco_serena}.
\end{figure}
\subsubsection{The only dynamic Rule : (Next)-rule}
\label{next_rule}

\bigskip 

In order to allow for a general but compact formulation of the $\nxt$-rule,  let us introduce first some notations. The goal is to define appropriate expansions for a premise  $E$ that is a set of assertions (disjunctively interpreted as usual) having the following general form:
$$\{ s_1 \vdash_\cgm [[A_1]] \nxt \varphi_1,...,  s_m \vdash_\cgm [[A_m]] \nxt \varphi_m, s'_{1} \vdash_\cgm\dlangle B_1 \drangle \nxt \psi_1,...,  s'_n \vdash_\cgm\dlangle B_n \drangle \nxt \psi_n  \}$$
\noindent
where  each $A_i$ and each $B_j$ is a coalition of agents, $n,m  \geq 0$ and $n+m \geq 1$.

Let us recall first that if $C$ is a coalition, $s$ is a state and $\sigma_C$ is a collective move of $C$ at $s$,  then: the notation $\ftrans(s, \sigma_\agents)$ denotes the state resulting when the global action $\sigma_\agents$ is played at state $s$, and    the notation $\mapAct{C}(s)$  denotes the set of all the $C$-moves available to coalition  $C$  at state $s$.  The expression $\outcome(s, \sigma_C)$ means the set of the states that are successors of $s$ when the coalition $C$ plays at $s$  the $C$-move  $\sigma_C$ and the other agents play no matter which move. Also, let us note $Co(C)$ the coalition complementary to $C$, namely $\agents \setminus C$. The connectives $AND$ and $\;  OR \; $ below are, respectively, a conjunction and a disjunction operator used as metalanguage operators to connect assertions, and they  have just a temporary role, to define abbreviations that are useful to formulate  the (Next)-rule.   Below we omit the subscript $\cgm$ in $\vdash_\cgm$.\\
\begin{itemize}
\item
For $1\leq i \leq m$,  $s$ a state, $\alpha \in \mapAct{A_i}(s)$  and $\delta$ an action  at state $s$ of  $co(A_i)$, that is a possible answer to $\alpha$ by  $A_i$'s opponents,  let us first note $\sigma(\alpha, \delta)$ the corresponding
global   action $\sigma_\agents$.
Then set
${Ans^{A_i,\varphi_i,s, \alpha, \delta}}$ to denote the  expression:
 $s"  \vdash \varphi_i$, where $s" = \ftrans(s, \sigma(\alpha, \delta) ) $.
This expression says that when $A_i$ plays $\alpha$ at state $s$ and its opponents answer with the action $\delta$, then $\varphi_i$ is true at the corresponding successor state.\\
\item 
Then  set
 $At\_least\_one\_ans^{A_i,\varphi_i,s, \alpha}$ to denote
 $OR_{\delta \in \mapAct{Co(A_i)}(s)}  \; Ans^{A_i,\varphi_i,s, \alpha, \delta}$.
 Intuitively, this last expression says that when $A_i$ plays $\alpha$ at state $s$  there is at least one answer of its opponents  that enforces the truth of  $\nxt \varphi_i$ at $s$.\\
Finally,  set $UnivSuccess(A_i, \varphi_i, s)$ to abbreviate the (meta) expression:\\
 $\;  AND \; _{\alpha \in    \mapAct{A_i}(s)}\;  At\_least\_one\_ans^{A_i,\varphi_i,s, \alpha}$.\\
 \textit{This conjunction 
 describes a sufficient and necessary condition for the truth at $s$ of the assertion $s \vdash [[A_i]] \varphi_i$}: it says that no matter 
  how  $A_i$  plays at $s$ there is at least 
 one answer  of its opponents that makes $\varphi_i$  true at the successor state.\\
\item 
For $1 \leq j \leq n$, $s$ a state, $\beta \in \mapAct{B_j}(s)$, 
set
 ${{Succes}^{B_j, \psi_j, s, \beta}}$ to be\\
  $ \;  AND \; _{s" \in \outcome(s, \beta)} \; s" \vdash \psi_j$. 
This conjunction   says  that  when coalition $B_j$ plays the action $\beta$ at $s$, $B_j$ succeeds  to assure $\psi_j$ at all  the reached $s$'s successors.
Then set $ExistSuccess^{B_j, \psi_j, s}$  to be
$\;  OR \; _{\beta \in \mapAct{B_j} (s)}{Succes}^{B_j,\psi_j,s,\beta}$\\
 \textit{This   describes a sufficient and necessary condition for the truth at $s$ of the assertion $\dlangle B_j \drangle \nxt \psi_j$}.\\
 \item 
  Finally, set $PreExpansion$ to be the expression 
 $$ (OR \; _{ 1 \leq i \leq m} UnivSuccess(A_i, \varphi_i, s)) \;  OR   \;  \;  (OR \; _{ 1 \leq j \leq n} ExistSuccess(A_i, \psi_i, s)) $$
 Let us observe that $E$ is true if and only if  $PreExpansion$ is true.\\
 
 Let's rewrite in conjunctive normal  form the expression $PreExpansion$ taking assertions as atoms, thereby obtaining a boolean combination  of assertions $a_p$:
$$(a_1 \;  OR \;  ... \;  OR \;  a_k) \;  AND \; ... \;  AND \;  (a_t \;  OR \;  ... \;  OR \;  a_q)$$ 
 \end{itemize}

  We are  now ready to formulate the (Next)-rule, shown in Figure\ref{next_rule}.
  \begin{figure}[h]
\vspace{-3mm}
\caption{Next-rule}
\label{next_rule}
\[\infer[(\nxt)]{
{a_1 ...  a_k||...||a_t  ...  a_q }        
}
{  s_1 \vdash_\cgm [[A_1]] \nxt G_1,...,  s_m \vdash_\cgm [[A_m]] \nxt G_m, s'_{1} \vdash_\cgm\dlangle B_1 \drangle \nxt F_1,...,  s'_n \vdash_\cgm\dlangle B_n \drangle \nxt F_n}
\]
\end{figure}
This rule has $n+m$ principal assertions and an \underline{empty context}.
\textit{This implies that it can be applied only when none of the static rules is any longer applicable}.
\scut{
The intuition behind the rule is that in order to validate an assertion
$ S \vdash \ll A \gg \bigcirc  F$
one verifies that there is at least one strategy that ensures $\bigcirc  F$, hence one checks that for at
least one collective $A$-action each obtained $S$'s  successor satisfies $F$, while to validate an assertion $S \vdash \neg \ll A \gg \bigcirc  G$ one checks that for each $A$-action there is at least one successor of $S$ thereby obtained where $G$ fails, i.e. $\neg F$ is satisfied.}

\subsection{Proofs in $\calcul$}
\begin{definition}[Candidate Proof]
\label{candidate}
Let $A$ be a set of assertions.
A \textit{candidate proof} $\cal P$ for $A$  is a directed rooted graph where vertices are labeled by sets of assertions that is built as follows:\\
1)  The root is labeled with $A$;\\
2) Let $A'$ be the label of  a vertex $v$. If $A'$ is $T$ or it is empty then $v$ is a leaf. Otherwise $v$  has children, and these last 
are labelled exactly  by the expansions obtained by applying an  inference rule to $A'$  as premise, with the proviso: there are no distinct vertices labeled by the same set of assertions along a path.
\end{definition}
The second condition allows for a finite representation of infinite cyclic paths and implies that a loop-check algorithm needs to be applied when paths are constructed, which we do in a depth-first manner, in order to ``point back" in case of cycles ; 
 hence the visited states of the path under construction must be memorized. 
In the sequel, we  identify vertices with the set of assertions  labeling them. 

The number of distinct sets of assertions that may occur in a candidate proof $\cal P$ is finite\footnote{This follows from the fact that the set of states in \cgm is finite and from the fact that only a finite number of  ``pseudo sub-formulae"  of a formula $\varphi$ occurring in a assertion at the root can occur in assertions in the graph. These formulae belong to  what is usually called the \textit{closure} of $\varphi$. }. Thus any infinite path $\pi$ in 
$\cal P$ must contain a cycle.  Say that a cycle is \textit{maximal} when it is not a proper sub-cycle of any cycle (\textit{i.e.} it is a strongly connected component).
Let us say, with an abuse of language, that $\pi$ \textit{ends with a maximal  cycle} whenever for some $j\geq 1$ and some $k > 1$  it has the form:
$$\pi =s_1,...,s_j,....,s_{j+k}, ....s_{j+2k},....,s_{j+3k}....$$
 where for any two distinct index $x$ and $y$  where $1 \leq x, y \leq j$  vertex $s_x$ and the vertex $s_j$ are different,  for any $q \geq 0$ and any $n>0$ we have $s_{j+q}= s_{j+nk+q}$ and $s_j,....,s_{j+k}$ is the period of a maximal cycle.
 \jeter{That is,  $\pi$ has the form:\\
 \begin{tikzpicture}
\node at (1,5.2) {\small{$s_1$}};
\node at (1,4) {\small{$s_{j}$}};``
\draw [->] [dotted, thick]  (1,4.9) --(1,4.2);
\draw [->] [dotted, thick]  (1,3.8) --(2,3)--(2.5, 3.5)--(1.3,4.2);
\end{tikzpicture} 

}

We call the vertex $s_j$ \textit{entry} of such a cycle.

\medskip

\jeter{
Let us observe that application of  the inference rules of $\calcul$ is non-deterministic, because, given any set of assertions $A=\{a_1,...,a_n\}$ labelling a vertex $s_j$ any $a_i$ can be chosen to be the principal assertion of an
inference rule.\footnote{With the exception of assertions $s \vdash_{\cgm} \psi$ where $\psi$ is a primitive successor formula, because the $(Next)$ rule requires the context to be empty.} However, we will always consider candidate proofs where the rule $(True)$ is applied with the highest priority. Say that a (finite or infinite) path $\pi = s_1, s_2, s_3,...$ is \textit{fair} whenever, given any assertion $a_i$, if $a_i $  has  been chosen as the principal formula to generate the vertex $s_{j+1}$ in the path and a second occurrence of  $a_i$ later in the path generates $s_k$, with $k> j+1$, then, whenever this is possible,  an assertion other than $a_i$ is chosen as the principal formula to generate a vertex  between $s_{j+1}$ and $s_k$. That is, each assertion is given a chance to be expanded.
\textit{We suppose that all the built paths are fair}. Hence, in
 just one case an assertion might not be analyzed:  the rule $True$  will be
  eagerly applied as soon as possible, since it allows for the immediate termination of the current path. The second highest priority is given to the rule $(False)$, which allows for the reduction of the size of the current label.
}

Say that a path formula is \textit{self-generating} when it has the form $\phi_1 \until \phi_2$ or  $\always \psi$,  or is a boolean combination  of formulae of these form.
Say that an assertion is \textit{potentially self-generating} when it has the form $s \vdash Q \; \Phi$ where $\Phi$ is a self-generating path formula.
Let us observe that only potentially self-generating assertions can be elements of the entry $E_j$ of a maximal cycle, because only a combination of $\gamma$-rules using fixed point equivalences and $(Next)$-rules can cause a loop on $E_j$.  

Among self-generating path formulae we distinguish  \textit{until formulae}.
 A formula $\phi_1 \until \phi_2$ is an until formula, a conjunction  of   until formulae is again
 an until formula, and the same for a disjunction of until formulae.
We call  \textit{until assertion} any assertion $s \vdash_\cgm Q \; \Phi$ where $\Phi$ is an until formula.

 \medskip
 
 Some cycles in an infinite path actually correspond to
failure to prove that what it means the set of assertions $A$ at the root is true, while other do not. This motivates the following definitions.
\begin{definition}[Failure Paths, Success Path, Proof]
\label{proof}
Let $\cal P$ be a candidate proof for a set of assertions $A$.\\
$\bullet$ A  finite  path $\pi$  is said to be a {finite} failure path when it ends with $\emptyset$,  and to be a 
{finite} success path when it ends with $T$.\\
$\bullet$ 
An infinite path $\pi$ is said to be   an \textit{infinite failure path} whenever it ends (``ends" in the sense made precise above) with a maximal cycle $$s_j,....,s_{j+k}, ....s_{j+2k},....,s_{j+3k}....$$ and 
all elements of the entry point $s_j$ are until assertions.
An  infinite path $\pi$ is said to be a \textit{infinite success path} when it is not a failure path.\\
$\bullet$ Let $\cal P$ be a candidate proof for $A$. It is a \textit{proof} of
$A$ whenever no path is a failure path.
Otherwise, $\cal P$  is said to fail.
\end{definition}
When the root $A$ of $\cal P$ is a singleton containing just an assertion   $a$, by an abuse of language  we will also speak of a candidate proof (or proof) for $a$.

The intuition behind the distinction made by Definition \ref{proof} between ``bad cycles", caused by until assertions, and ``good cycles", where there is a loop on assertions of the form
$s \vdash \always \phi$ is semantical: a cycle on $s \vdash  \dlangle A \drangle \phi_1 \until \phi_2$, for instance, shows an infinite procrastination of the realization of the eventuality 
$\phi_2$, while a cycle on $s \vdash  \dlangle A \drangle \always  \phi_1$ is not harmful: if the CGM \cgm has a cycle having period $s=s_1,...,s_n$, $s_1$ being a successor of $s_n$,
the falsity of this box-assertion would be witnessed by the falsity of $\phi_1$ at some state $s_i$ where $1 \leq i \leq n$, while the corresponding cycle in $\cal P$  just shows that 
$\phi_1$ must be true everywhere in order for $s \vdash  \dlangle A \drangle \always  \phi_1$ to hold for \cgm.








\section{Some examples}
We give below some simple exemples  that should help the reader  to better grasp the intuition behind Definition \ref{proof}.
\subsubsection{Failing $\always$ assertions.}$\;$\\

First, observe that by semantical reasons 
an assertion of the form $s\vdash_\cgm \diaA \always \Phi$  is false  whenever for each strategy $\stratA$ for the coalition $A$ there is at least one  finite initial segment of a run in $\cgm$ compliant with $\stratA$, say  
$\lambda$ $=$
$\lambda_1, ...., \lambda_n$, $n \geq 1$,  such that 
$\lambda_1=s$,  $\lambda_n \not  \models\Phi $ (and $n$ can be taken to be the least natural satisfying the property, possibly $n=1$).
Let $\cal P$  any candidate proof  for  $s\vdash_\cgm \diaA \always \Phi$. There will always at least a finite failure path witnessing this semantical feature and
\scut{\scomment{Ehm, jusqu'à la c'est juste une conjecture, car ce que je dis est une partie de la soundness... De plus, ma phrase est bien vague et je le sais.}}
\scut{Consider those paths in $\cal P$ where the expansions for the rule $(\gamma)$ applied to $\always$ or $\until$ assertions
 (the only rule instances that can be "responsible" for cycles in $\cal P$) are chosen with the following criterion: ``\textit{whenever possible, choose always a false expansion not producing a copy of the expanded assertion}". That is, if, for instance,  the  $(\gamma)$-rule  is applied to   $A, s \vdash_\cgm\diaA \always q$ and produces 
the expansions $A, s'  \vdash_\cgm q$ and $A, s''  \vdash_\cgm \diaA \nxt \diaA \always q$, and both these expansions are false, the expansion $A, s'  \vdash_\cgm q$ is chosen. Taking expansions for the rule $(\gamma)$ applied 
to any $\until$ or $\always$ assertion to be ordered so that expansions not containing a copy of the expanded assertion are at the left of expansions 
containing such a copy, a strategy of path construction of this sort  can be briefly formulated as ``\textit{whenever possible, choose always the leftist expansion that is false}".
These paths  will always be finite failure paths and willl describe such failing runs $\lambda$. Let us call a strategy for building paths in a proof structure satisfying the above criterion  \textit{leftist falsifying path construction strategy (LFPS)}.}
there will be a finite failure path describing such a failure.
 \begin{example}
 \label{failing_always_exa}
 We illustrate the above observation with a very simple example, where in the CGM \cgm $\;$  we have just one agent, 1, that can play only one action:
 
 \medskip

\begin{tikzpicture}
\draw [thick] (1,5) circle [radius=0.3];
\draw  [thick] (3,5) circle [radius=0.3];
\node at (1,5) {\small{$s1$}};
\node at (3,5) {\small{$s2$}};
\node at (1,4.3) {\small{$\{q\}$}};``
\node at (3,4.3) {\small{$\emptyset$}};
\draw [->] [thick]  (1.5,5.2) --(2.6,5.2);
\draw [->] [thick]  (2.6,4.8) --(1.5,4.8);

\end{tikzpicture} 
 
The atom $q$ is true only at $s1$ and  $s1 \not \models \dlangle 1 \drangle \always q$. 
An initial segment of the unique  path in this model, namely  $\lambda \; = \; s1, s2$ suffices to witness this.

Below we exhibit a failing proof structure for  
$s_1 \vdash_\cgm \dlangle 1  \drangle \always q$
showing that such  assertion is false.  For the sake of clarity, we enumerate vertices.

\begin{tikzpicture}
\node at (4,7) {{$1: s_1 \vdash_\cgm \dlangle 1  \drangle \always q$}};
\node at (2,5.5) {{$2: s_1 \vdash_\cgm q$}};
\node at (6,5.5) {{$3: s_1 \vdash_\cgm   \dlangle 1  \drangle \nxt \dlangle 1  \drangle \always q$}};
\node at (4,6.3) {\small{$(\gamma)$}};
\draw [->] [thick]  (4,6.8) --(2,5.8);
\draw [->] [thick]  (4,6.8) --(6,5.8);
\node at (2,3.6) {{$4: \top$}};
\node at (6,3.6) {{$5: s_2 \vdash_\cgm \dlangle 1  \drangle \always q$}};
\draw [->] [thick]  (2,5.3) --(2,3.8);
\draw [->] [thick]  (6,5.3) --(6,3.8);
\node at (2.4,5) {{$(True)$}};
\node at (6.4,5) {\small{$ (Next)$}};
\node at (4,1.6) {{$6: s_2 \vdash_\cgm  q$}};
\node at (9,1.6) {{$7: s_2 \vdash_\cgm   \dlangle 1  \drangle \nxt \dlangle 1  \drangle \always q$}};
\draw [->] [thick]  (6,3.3) --(4,1.9);
\draw [->] [thick]  (6,3.3) --(8,1.9);
\node at (6,2.5) {\small{$(\gamma)$}};
\draw [->] [thick]  (4,1.4) --(4,0.2);
\node at (4,0) {{$8: \emptyset$}};
\node at (4.5,1) {{$(False)$}};
\draw [->] [thick]  (9,1.8)--(11,4.5)--(9.5,5.5)-- (5.6,7);
\node at (9.5,4.5) {{$\small(Next)$}};

\end{tikzpicture} 
 
In such a proof structure the path consisting of the vertices  1, 3, 5, 6, 8  is a finite failure path and describes the finite sequence of states in the model  that suffices to witness the falsity of  $s_1 \vdash_\cgm \dlangle 1  \drangle \always q$, namely $s1, s2$.

 \end{example}


\jeter{
\subsubsection{Failing  $\until$ assertions.}$\;$\\

Things are different for the falsity of an assertion of the form $s \vdash_\cgm \dlangle A \drangle  \Phi_1 \until \Phi_2$, where possibly no finite failure path in
a candidate proof for $s \vdash\cgm \dlangle A \drangle  \Phi_1 \until \Phi_2$ will show the falsity of the root, as witnesses by the following  very simple example. 

\begin{example}
\label{quattro}
Consider a CGM $\cgm$ 
where we have an unique agent, 1, an unique action $a$, just one state $s$, where $p$ is true but $q$ is false, 
and a loop on $s$. Let us consider the false assertion $s \vdash_\cgm \dlangle 1 \drangle p \until q$.
In the only possible candidate  proof  for $s \vdash_\cgm p \until q$ the root  has two children, generated by an application of the $\gamma$-rule: 2, labelled by $s \vdash_\cgm p,q$, that leads immediately to $T$, and 3, labelled by 
by the set $s \vdash_\cgm q, \;\;  s  \vdash_\cgm \langle 1 \drangle \nxt \dlangle 1 \drangle  p \until q$. 
Vertex 3 generates again the same label as 1 via the rule $False$ (applied to $1 \vdash_\cgm q$) followed by the rule $Next$. Hence we have a path that is infinite and fails. The involved cycle 
shows the impossibility of realizing the eventuality.

\jeter{

\begin{tikzpicture}
\node at (5, 11) {{$1 : s \vdash \dlangle 1 \drangle p \until q$}};
\node at (1,8) {{$2 : s \vdash   p, \; s \vdash q$}};
\node at (8,8) {{$3 :  s \vdash q, \; 
s \vdash \dlangle 1 \drangle \nxt  \dlangle 1 \drangle  p \until q$}};
\node at (5,10) {\small{$(\gamma)$}};
\draw [->] [thick]  (5,10.8) --(1,8.3);
\draw [->] [thick]  (5,10.8) --(8.5,8.3);
\node at (1,6) {{$4 : T$}};
\draw [->] [thick]  (1,7.8) --(1,6.2);
\node at (1.5,7) {{$(True)$}};
\draw [->] [thick]  (8,7.8) --(8,6.2);
\node at (9,7) {{$(False)$}};
\node at (8,6) {{$5 : s \vdash \dlangle 1 \drangle \nxt  p \until q$}};
\draw [->] [thick]  (7.9,6.2) --(12,7.5)--(7,10)--(5.2,10.8);
\node at (10,9) {{$(Next)$}};

\end{tikzpicture}
\medskip

}

On the other hand, considering always the same CGM \cgm, the only possible candidate proof $\cal P$ for the true
assertion $s \vdash_\cgm \dlangle 1 \drangle \always \neg q$ has just two paths: a finite one, ending with $T$, that is a finite success, and the cyclic one:\\
$s \vdash_\cgm \dlangle 1 \drangle \always \neg q, \; s \vdash_\cgm \dlangle 1 \drangle \nxt \dlangle 1 \drangle \always \neg q, \; s \vdash_\cgm \dlangle 1 \drangle \always \neg q, ...$\\
This cycle does not prevent $s \vdash_\cgm \dlangle 1 \drangle \always \neg q$ from being true, and actually the corresponding path is not an infinite failure path: $\cal P$ is a true proof of the root.
\end{example}

}

\subsubsection{More examples.}$\;$\\

The next example is very simple, but it suffices to show why a maximal cycle induces a failure  only when 
\textit{all} the elements of its entry point are until formulae.

\begin{example}
Let us consider again the same simple GCM of Example \ref{quattro}. The assertion  $ a = s \vdash ( (\dlangle 1 \drangle  \always p) \vee ( \dlangle 1 \drangle  \event \neg p))$ is trivially true. Let us consider the following candidate proof:

\begin{tikzpicture}

\node at (5, 11) {{$1 : a $}};
\draw [->] [thick]  (5,10.8) --(5,10.2);

\node at (5, 10) {{$2  :  \; s \vdash \dlangle 1 \drangle  \always p$,  $ s \vdash  \dlangle 1 \drangle  \event \neg p$}};
\draw [->] [thick]  (5,9.8) --(2,9.2);

\node at (2, 9){{$3 : s \vdash p,  \; s \vdash    \dlangle 1 \drangle  \event \neg p$}};
\draw [->] [thick]  (5,9.8) --(9,9.2);

\node at (9, 9){{$5 : s \vdash \dlangle 1 \drangle \nxt   \dlangle 1 \drangle  \always p,  \; s \vdash    \dlangle 1 \drangle  \event \neg p$}};
\draw [->] [thick]  (1,8.8) --(1,8.2);

\draw [->] [thick]  (8,8.8) --(8,8.2);

\node at (1,8) {{$4 : True$}};
\node at (8,8) {{$6 :   s \vdash \dlangle 1 \drangle \nxt   \dlangle 1 \drangle  \always p,  \;  \neg p,  \; s \vdash \dlangle 1 \drangle \nxt     \dlangle 1 \drangle  \event \neg p$}};
\node at (8,7) {{$7 :   s \vdash \dlangle 1 \drangle \nxt   \dlangle 1 \drangle  \always p,   \; s \vdash \dlangle 1 \drangle \nxt     \dlangle 1 \drangle  \event \neg p$}};
\draw [->] [thick]  (8,7.8) --(8,7.2);

\draw [->] [thick]  (11.5,7) --(13,8.5)--(9,10)--(5.2,10.8);

\end{tikzpicture}

We see that even if vertex 1, the entry of the cycle, contains an until assertion, that is thereby generated over and over, the cycle does not induce a failure path; indeed the assertion
$\dlangle 1 \drangle  \always p $ also belongs to such a vertex, and its truth causes the true of the  root of the candidate proof. Here we have a proof of assertion $a$.

\end{example}






The next example is more complex, although the involved formulae are all Vanilla-ATL formulae, as in the previous ones.

\begin{example}
Consider a CGM with three states,$ A$,$ B$  and $C$, and two agents, 1 and 2. Agent 1 can play action $a$ at each state and agent $B$ can play action $b$ at each state ; moreover, $B$ can also play $b'$ at state $A$. Variable, $p$  is true only at $C$ and variable $r$ is false everywhere. The transitions are:\\

\begin{tikzpicture}

\node at (1, 4) {\small{$A$}};
\draw [thick] (1,4) circle [radius=0.3];

\node at (5, 5) {\small{$B$}};
\draw [thick] (5,5) circle [radius=0.3];

\node at (5, 3) {\small{$C$}};
\draw [thick] (5,3) circle [radius=0.3];

\draw [->] [thick]  (1.3,4) --(4.7,5);

\draw [->] [thick]  (1.3,4) --(4.7,3);

\draw [->] [thick]  (4.8,5.2) --(1.2,4.2);

\draw [->] [thick]  (4.7,2.8)--(1.3,3.8);

\node at (2.5, 4.9) {\small{$\langle a, b \rangle$}};

\node at (4, 4.5) {\small{$\langle a, b \rangle$}};

\node at (3.5, 3.7) {\small{$\langle a, b' \rangle$}};

\node at (2.5, 3) {\small{$\langle a, b \rangle$}};


\end{tikzpicture}

The assertion $A \vdash \dlangle 2 \drangle (\dlangle 1 \drangle p) \until r$ is false at $A$.  A corresponding candidate proof has 46 vertices and we do not exhibit it completely here.
We show, however, its general structure. We number vertices depth-first, we use $\rightarrow$ also  to draw a segment of a  path whose length may be greater than 1,  and we abbreviate the formula 
$\dlangle 2 \drangle (\dlangle 1 \drangle p) \until r$ with $U$ (as ``until"),  the formula $\dlangle 1 \drangle \event p$ by $D$. 

\begin{tikzpicture}

\node at (5, 11) {{$1 :  A \vdash U$}};
\draw [->] [thick]  (5,10.8) --(2,10.2);
\node at (2.5, 10) {{$3 :  A \vdash D $}};
\draw [->] [thick]  (2,9.8) --(2,9.2);
\node at (2, 9) {{$5$}};
\draw [->] [thick]  (2,8.8) --(1,8.2);
\node at (1, 8) {{$8: B \vdash \dlangle 1 \drangle  \nxt D $}};
\draw [->] [thick]  (2,8.8) --(3.8,8.2);
\draw [->] [thick]  (1, 7.7)--(0.7, 8)--(0.5, 9.3)--(2.5,9.8);

\node at (3.5, 8) {{$11: T $}};

\draw [->] [thick]  (5,10.8) --(8,10.2);
\node at (8.3, 10) {{$13$}};
\draw [->] [thick]  (8.3,9.8) --(7,9.2);

\draw [->] [thick]  (8.3,9.8) --(11,9.4);

\draw [->] [thick]  (7,8.8) --(5.8,8.2);

\node at (7, 9) {{$14$}};

\draw [->] [thick]  (7,8.8) --(8.7,8.2);
\node at (11, 9.2) {{$27$}};
\draw [->] [thick]  (11,9.1) --(11,8.6);

\draw [->] [thick]  (11, 9.2)--(12.5, 8.6);

\node at (11, 8.4) {{$31: T$}};

\node at (5.8, 8) {{$16: B \vdash D$}};
\draw [->] [thick]  (5.8,7.8) --(5.8,7.2);

\draw [->] [thick]  (5.6,7.2)--(5.6,7.8);

\node at (5.8, 7) {{$21: A \vdash \langle 1 \rangle  \nxt D$}};
\draw [->] [thick]  (5.8,6.8) --(6.3,6.2);

\node at (6.4, 6) {{$24: T$}};

\node at (8.7, 8) {{$26: B \vdash  \dlangle 2 \drangle \nxt U$}};
\draw [->] [thick]  (8.7, 8)--(5.7, 9)--(5,10.6);

\node at (12.5, 8.4) {{$34$}};

\draw [->] [thick]  (12.5, 8.2)--(12, 7.6);

\draw [->] [thick]  (12.5, 8.2)--(13, 7.6);

\node at (11.8, 7.4) {{$36: B \vdash D$}};

\draw [->] [thick]  (11.8, 7.2)--(10, 6.6);

\draw [->] [thick](9, 6.6) --  (10.8, 7.2);

\node at (10, 6.4) {{$41: A \vdash \dlangle 1 \drangle \nxt  D$}};

\draw [->] [thick]  (9.9, 6.2)--(10.5, 5.8);

\node at (10.5, 5.6) {{$42: T$}};

\node at (14.4, 7.4) {{$46: B \vdash \langle 2 \rangle \nxt U$}};

\draw [->] [thick](14.5, 7.6)--(15,9.9)-- (5, 10.8);

\end{tikzpicture}

The above graph is not a proof, and for many reasons, since there are several maximal cycles, and the entry point of each of them contains only until formulae.
\end{example}

Our last example involves a formula that, although very simple,  is not a Vanilla \ATL formula.

\begin{example}
\label{esempio_Piu}
 Let us consider again the CGM of Example  \ref{quattro} and the false assertion
$s \vdash \dlangle 1  \drangle (\event q \wedge \event p)$.
It is tedious but easy to check that the corresponding proof candidate contains an infinite failure path cycling on $s \vdash \dlangle 1  \drangle (\event q \wedge \event p), s  \vdash \event q $.

\end{example}

\section{Properties of the calculus}

\subsection{Some basic properties of $\calcul$}

Let us recall that we consider only candidate proofs whose paths are fair. Thus, for any path $\lambda$, 
 each assertion $a$ occurring in the label of a vertex  has a chance to be expanded, unless  the rule $True$ causes $\lambda$ to halt before
 analyzing $a$.

The very definition of the rules immediately implies the first of the two properties stated by the following lemma, which in its turn implies the second property:
\jeter{

\begin{lemma}
\label{rev_prop_lemma}
$\;$\\
\begin{enumerate}
\item

Each  rule of $\calcul$, but $(True)$, is reversible in the sense that, for any given CGM  $\cgm$,  its premise is  true if  and only if  the conjunction of its expansions is true. 

\item
\label{permu}
Let $R$ et $R'$  be any two rule instances  in  $\calcul$, excepted for $(\nxt)$ and $(True)$. Their order of application does not matter, in the following sense. 
Let $\cal P$ be a candidate proof for  $A= a_1, ..., a_n$ 
where first $R$ is applied to $a_i$, then $R'$ is applied to $a_j$, $i \not = j$, thereby getting conclusions $a'i_1,...,a'_m$. 
Then there is also a candidate proof ${\cal P}' $  for  $A$
where first $R'$ is applied to $a_j$, then $R$ is applied to $a_i$, getting the same conclusions  $a'i_1,...,a'_m$.

\end{enumerate}
\end{lemma}
}

\jeter{
\begin{definition}
\label{leaf_cycle}
Given a candidate proof $\cal P$ say that a vertex is a \textit{quasi-leaf} when either it is the premise of the rule $(True)$, or it is the premise of $(False)$ and its conclusion is empty,
or else it is the   entry point of a maximal cycle. 
\end{definition}

Lemma \ref{rev_prop_lemma}
 immediately implies:

\begin{proposition}
\label{root_leaves_property}
Given any candidate proof $\cal P$, the label of its root is true with respect to a given CGM  $\cgm$ if and only if all the labels of its quasi-leaves are so.
\label{rev_prop}

\end{proposition}
}

\begin{definition}[closure]
The closure $cl(\varphi)$ of an $\ATLp$ state formula $\varphi$  is the least set of \ATLp formulae such that $\top$ and $\varphi$ are in $cl(\varphi)$ and $cl(\varphi)$ is closed under taking successor, $\alpha$, $\beta$ and $\gamma_d$ components of $\varphi$. If $\Gamma$ is a set of state \ATLp formulae we set:
$$cl(\Gamma)= \{\psi \;\mid\; \psi \in \Gamma\}$$
\end{definition}

An inspection of the rules of $\calcul$ immediately shows:

\begin{lemma}
\label{closure}
For 
any candidate proof $\cal P$ for a set of assertions $A=\{s_1 \vdash_\cgm \varphi_1, ... , s_n \vdash_\cgm \varphi_n\}$, any assertion $s \vdash_\cgm \psi$ occurring in $\cal P$
is such that $s$ is a state of $\cgm$ and $\psi \in cl(\{\varphi_1,..,\varphi_n\})$. 
\end{lemma}

The next lemma describes a property of proofs of until assertions.
\begin{lemma}
\label{speriamo}
Let $\cgm$ be a CGM, let $\cal P$ be any candidate proof.
\begin{enumerate}
\item
\label{existQ}
Let $u=s_1 \vdash_{\cgm} \dlangle A \drangle \phi_1 \until \phi_2$ be a true until assertion.

Let $\stratA$ be any strategy for the coalition $A$ witnessing the truth of such an assertion in \cgm, that is, such that  any branch $\pi$ of  the tree ${\cal T}_{\cgm, \stratA}$
of the paths in the CGM \cgm that are compliant with 
$\stratA$ and are
 rooted at $s_1$  has the property: \\
it has the form  $s_1, s_2, ...., s_k,....$ for some
$k \geq 1$,  $s_k \models \phi_2$, and $\phi_1$ is true at each $s_j$ in $\{s_1,...,s_{k-1}\}$.\\
 Without any loss of generality we can suppose that $k$ is the least index having such a property.

Let $\pi$ be any branch in  ${\cal T}_{\cgm, \stratA}$.

Suppose that  $\cal P$ contains, along one of its paths,  some vertex $r$ having  the until assertion $u$ in its label, and expanded by a $\gamma$ rule having such an assertion has its principal assertion. Then the sub-graph of $\cal P$ rooted at $r$ starts with the following tree ${\cal T}_u$ (omitting the subscript $\cgm)$, where a dotted arc represents a finite sequence of applications of static rules:

 \begin{tikzpicture}
\node at (3.5,11) {\small{$s_1  \vdash \dlangle A \drangle \phi_1 \until \phi_2,...$}};
\draw [->] [thick]  (3.7,10.8) --(1,10.2);
\node at (1,10) {\small{$s_1 \vdash \phi_1, s_1 \vdash \phi_2,...$}};
\draw [->] [thick]  (3.8,10.8) --(7,10.2);
\node at (7,10) {\small{$s_1 \vdash \phi_2, \; s_1  \vdash \dlangle A \drangle  \nxt  \dlangle A \drangle \phi_1 \until \phi_2,...$}};
\draw [->] [dotted,thick]  (7,9.8) --(7,8.2);
\node at (9.5,8) {\small{$ s_1  \vdash \dlangle A \drangle  \nxt  \dlangle A \drangle \phi_1 \until \phi_2$, only  successor-formulae}};
\node at (9,7.5) {\small(Next rule)};
\draw [->] [thick]  (10,7.8) --(10,7.2);
\node at (9,7) {\small{$s_2  \vdash \dlangle A \drangle \phi_1 \until \phi_2,...$}};
\draw [->] [thick]  (9,6.8) --(7,6.2);
\node at (7,6) {\small{$s_2 \vdash \phi_1, s_2 \vdash \phi_2,...$}};
\draw [->] [thick]  (9,6.8) --(12,6.2);
\node at (12,6) {\small{$s_2 \vdash \phi_2, \; s_2  \vdash \dlangle A \drangle  \nxt  \dlangle A \drangle \phi_1 \until \phi_2,...$}};
\draw [->] [dotted,thick]  (12,5.8) --(12,4.2);
\node at (11.5,4) {\small{$s_{k-1} \vdash  \dlangle A \drangle  \nxt  \dlangle A \drangle \phi_1 \until \phi_2,...$,  only  successor-formulae}};
\draw [->] [thick]  (11,3.8) --(11,3.2);
\node at (12,3.5) {\small(Next rule)};
\node at (11,3) {\small{$s_k \vdash \dlangle A \drangle \phi_1 \until \phi_2,...$}};
\draw [->] [thick]  (11,2.8) --(9,2.2);

\node at (9,2) {\small{$s_k \vdash \phi_1, s_k \vdash \phi_2,...$}};

\draw [->] [thick]  (11,2.8) --(12,2.2); 
\node at (12,2) {\small{$s_k \vdash \phi_2,...$}};

\end{tikzpicture} 

\noindent
where:
\begin{itemize}
\item
For each $1 \leq i \leq k-1$, the assertion $s_i \vdash \phi_1$ is true
\item 
The assertion $ s_{k} \vdash \phi_2$ is true.
\end{itemize}

\medskip

We will say that any path in $\cal P$ leading to $r$ and continuing, after $r$, as shown by  some branch of the above tree  ${\cal T}_u$ (and no matter how after)
 represents $\pi$.
 
 \medskip

 In this sense, whenever $\cal P$ contains   some vertex $r$ having  the until assertion $u$ in its label,  expanded by a $\gamma$ rule having such an assertion has its principal assertion, \textit{some tree ${\cal T}_u$ in $\cal P$ represents   ${\cal T}_{\cgm, \stratA}$ and any branch in 
 in ${\cal T}_u$ partially represents a branch $\pi$ in ${\cal T}_{\cgm, \stratA}$}.\\

 \item
 \label{univQ}

 Let $u=s_1 \vdash_{\cgm} [[A]] \phi_1 \until \phi_2$ be a true until assertion. 

Let ${\stratA}^c$ be any co-strategy for the coalition $A$ witnessing the truth of such an assertion in \cgm, that is, such that  any branch $\pi$ of  the tree ${\cal T}_{\cgm, {\stratA}^c}$
of the paths in the CGM \cgm that are compliant with ${\stratA}^c$ 
and are
 rooted at $s_1$  has the property: \\
it has the form  $s_1, s_2, ...., s_k,....$ for some
$k \geq 1$,  $s_k \models \phi_2$, and $\phi_1$ is true at each $s_j$ in $\{s_1,...,s_{k-1}\}$.\\

 Again,  whenever $\cal P$ contains   some vertex $r$ having  the until assertion $u$ in its label, expanded by a $\gamma$ rule having such an assertion has its principal assertion, \textit{some tree ${\cal T}_u$ in $\cal P$ represents   ${\cal T}_{\cgm, {\stratA}^c}$
  and any branch in 
 in ${\cal T}_u$ partially represents a branch $\pi$ in ${\cal T}_{\cgm, \stratA}$}, as in the previous item.

\end{enumerate}


\end{lemma}

\noindent
\textit{Proof}\\
The truth of item 1 is an immediate consequence of the definition of the expansion rules, that closely follows $\ATLp$ semantics. The reasoning for item 2 is exactly the
same, modulo using the notion of co-strategy rather than the notion of strategy.\\

\noindent
$\Box$\\

It is  worthwhile observing that the same path in a candidate proof can  represent different paths in the model, because a vertex in a proof is a labeled by assertions involving several states.

\bigskip

\begin{theorem}[Soundness and Completeness]
\label{comple_th}
The calculus $\calcul$ is sound and complete:  $\cal P$ is a proof for  an assertion $a$ if and only if $a$ is true.
\end{theorem}

\noindent
\subsubsection*{Proof of Theorem \ref{comple_th} : from left to right implication.}

What we need prove here is: \textit{If  $\cal P$ is a  proof for $a$ then   $a$ is true}.

\bigskip

Let us first recall that any infinite path $\pi$  ``ends" with a maximal cycle. That is, for some $j$ and $k$ $\geq 0$, 
$$\pi= E_1,...,E_j,....,E_{j+k}, ....E_{j+2k},....,E_{j+3k}....$$
 where each $E_i$ is a set of assertions, and:
 \begin{itemize}
 \item $E_j$ is the f entry in the cycle;
 \item For $1\leq i,t \leq j$ such that $ i \not = t$ we have $E_t \not = E_j$; 
 \item  For any $q \geq 0$ and any $n>0$ we have $E_{j+q}= E_{j+nk+q}$, \textit{i.e.} the sub-path $E_j,....,E_{j+k}$ is the period of  a maximal  cycle.
 \end{itemize}
Hence $\pi$  has the form:

\begin{tikzpicture}
\node at (1,5.2) {\small{$E_1$}};
\node at (1,4) {\small{$E_{j}$}};``
\draw [->] [dotted, thick]  (1,4.9) --(1,4.2);
\draw [->] [dotted, thick]  (1,3.8) --(2,3)--(2.5, 3.5)--(1.3,4.2);
\end{tikzpicture} 

\noindent

By definition, in order for $\pi$ to be successful, not all the elements of $E_j$ are until assertions.

It is immediate to see that:\\
Fact a) If a sub-graph in a proof $\cal P$ contains only finite paths then the set of assertions at its root is true (\textit{i.e}. at least one oof its elements is true), because by definition of proof any of its paths  $\pi$ is successful, hence ends with $T$, and Proposition \ref{root_leaves_property} holds.

\medskip

Hence what we prove here is:\\
Fact b)  If  $G_{E_{j}}$ is the subgraph of a proof $\cal P$  rooted at the first entry point $E_j$ of a maximal cycle then the set of assertions $E_j$ is true. 

Once proved this, the result will follow by Proposition \ref{root_leaves_property}.


In order to prove (b) we use an appropriate notion of modal depth, which  takes into account only the nesting of the $\always$ modal operators.
The \textit{box modal depth} of a state formula $\phi$, noted $BMD(\phi)$,  is defined by induction on path-formulae of $\ATLp$ in negation normal form (state formulae being a particular case of path formulae) as it follows:
\begin{itemize}
\item If $\Phi$ is a literal $l$ then  $BMD(\Phi)=0$
\item  If $\Phi$ is  $\psi_1 \; c \; \psi_2$ then $BMD(\Phi)=   Max\{BMD(\Psi_1),B MD(\Psi_2)\}$ for $c \in \{\wedge, \vee\}$
\item If $\Phi$ is $Q \; \Psi$ then $BMD(\Phi)= BMD(\Psi)$ for $Q \in \{\diaA, [[A]]\}$ and $A$ a coalition 
\item If $\Phi$ is $\nxt \psi$  then  $BMD(\Phi)= MD(\psi)$
\item If $\Phi$ is $\always \psi$ then $BMD(\Phi)=B MD(\always \psi)+1$
\item If $\Phi$ is $ \psi_1 \until \psi_2$ then $BMD(\Phi) = Max\{BMD\psi_1), BMD(\psi_2)\}$ 
\end{itemize}

 If $S$ is any set of assertions then its
 modal depth  is  the maximal value of the box modal depths of its elements. 
 
 Thus, let   $G_{E_{j}}$ be the subgraph of a proof $\cal P$  rooted at $E_j$.  Let $E_j$ be $B \cup U_1$ where $U_1$ is the set of all the until assertions in $E_j$
 and $B$ is the set of potentially  self-generating assertions that are not until formulae.

 We  prove (b) by induction on $BMD(B)$.

\noindent
\textit{Basis}. $BMD(B)= 1$.  
In this case  any element of $B$ has the form $s_1 \vdash Q \;  \always \varphi$ for some state $s_1$ of the considered CGM, where  $\varphi $ is a state formula
 having  modal depth 0, so no occurrence of $\always$. 
For the sake of readability we reason here for the case where $B$ has just an element, but the reasoning is essentially the same in the general case (although the corresponding figure
is more difficult to draw and read).
The sub-graph $G$ of $\cal P$ will have the form:

\medskip

 \begin{tikzpicture}
\node at (3.5,11) {\small{$s_1  \vdash \dlangle A\drangle \always \psi, U_1$}};
\draw [->] [thick]  (3.7,10.8) --(1,10.2);
\node at (1,10) {\small{$s_1 \vdash \psi, U_1$}};
\draw [->] [thick]  (3.8,10.8) --(7,10.2);
\node at (7,10) {\small{$s_1  \vdash \dlangle A \drangle  \nxt  \dlangle A \drangle \always \psi, U_1$}};
\draw [->] [dotted,thick]  (7,9.8) --(7,8.2);
\node at (9.5,8) {\small{$ s_1  \vdash \dlangle A \drangle  \nxt  \dlangle A \drangle \always \psi, S_1\; $  only  successor-formulae}};
\node at (7,7.5) {\small(Next rule)};
\draw [->] [thick]  (8,7.8) --(8,7.2);
\draw [->] [thick]  (8,7.8) --(12,7.2);
\draw  [dotted,thick]  (8,7.8) --(13,7.6);
\node at (9,7) {\small{$s_2  \vdash \dlangle A \drangle \always \psi, U_2$}};
\node at (12,7) {\small{$s'_2  \vdash \dlangle A \drangle \always \psi, U'_2 \; \; ...$}};
\draw  [dotted, thick]  (12,6.7) --(13.5,6.3);
\draw [->] [thick]  (9,6.8) --(7,6.3);
\node at (7,6) {\small{$s_2 \vdash \psi, U_2$}};
\draw [->] [thick]  (9,6.8) --(12,6.2);
\node at (12,6) {\small{$s_2  \vdash \dlangle A \drangle  \nxt  \dlangle A \drangle \always \psi, U_2$}};
\draw [->] [dotted,thick]  (11,5.8) --(11,4.3);
\node at (11.5,4) {\small{$s_{k-1} \vdash  \dlangle A \drangle  \nxt  \dlangle A \drangle \always \psi, S_{k-1} \; $ only  successor-formulae}};
\draw [->] [thick]  (11,3.8) --(11,3.2);
\node at (12,3.5) {\small(Next rule)};
\node at (12,3) {\small{$s_k \vdash \dlangle A \drangle \always \psi, U_k \; \; ...$}};
\draw [->] [thick]  (11,2.8) --(9,2.2);

\node at (8,2) {\small{$s_k \vdash \psi, U_k$}};

\draw [->] [thick]  (11,2.8) --(11.6,2.3); 
\node at (12,2) {\small{$s_k \vdash \dlangle A \drangle  \nxt  \dlangle A \drangle \always \psi, S_k \; $  }};
\draw [->] [thick]   (12.3,2.2) --(6.5,5)--(3.5, 10.7);
\end{tikzpicture}

\bigskip

\noindent
Without loss of generality we have supposed that the root $s_1  \vdash \dlangle A\drangle \always \psi, U$ has been expanded via a $\gamma$-rule
having $s_1  \vdash \dlangle A\drangle \always \psi$ as main assertion (see the second item of Lemma \ref{rev_prop_lemma}).


For each  formula  $ \phi_1 \until \phi_2$ generated   by an element   $s \vdash \xi$ in $U_1$  in the path which goes from  $s_1  \vdash \dlangle A \drangle  \nxt  \dlangle A \drangle \always \psi, U_1$ to $s_1  \vdash \dlangle A \drangle  \nxt  \dlangle A \drangle \always \psi, S_1$ (dotted in the figure), 
 a  vertex  $s  \vdash  \phi_2,  s_1 \vdash Q  \nxt Q \; \phi_1 \until \phi_2,...$ is generated in that path.
The until formulae of $U_1$ are  generated over and over in the cycle shown in the figure.
  In order to get the cycle, an application of the Next-rule is needed, producing  the vertices $s_2  \vdash \dlangle A \drangle \always \psi, U_2$, $s'_2  \vdash \dlangle A \drangle \always \psi, U'_2 ...$, 
where $s_2$, $s'_2$,... are the successors of $s_1$ in the CGM $\cgm$. But the (Next)-rule can be applied only when the assertions in its premise 
are all successor formulae.
Therefore
  $s \vdash  \phi_2$ must have disappeared when (Next) is applied.  Only an application of the (False) rule can cause such a cancellation of
  $s \vdash  \phi_2$. Then  the assertions   $s' \vdash \phi_2$, must all be false, and $U_1$ is false.

 The set of assertions $\{s_1 \vdash \psi, U_1\}$ has modal depth 0, so $\psi$ does contain any occurrence of $\always$.   Therefore if the sub-graph rooted at it contained a cycle,  this
 might   be caused only by until-assertion, and it would be a bad cycle,  contrarily to our hypothesis that $\cal P$ is a proof. Hence such a sub-graph contains only
 finite paths. By the fact (a) above $\{s_1 \vdash \psi, U_1\}$ is true, which implies that $s_1 \vdash \psi$ is true, since $U_1$ is false..

 By the same reasoning we get that $s_2  \vdash \psi, U_2$, ...., $s_k  \vdash \psi, U_k$
 are true, and similarly for the analogous sets corresponding to any other $s'$ that is a successor of $s_1$, to any successor of such a $s'$ etc (for readability the figure 
 does not show them) and each time it is the corresponding assertion $s \vdash \psi$ the cause of their truth.  Hence for each  $l$, $1 \leq l \leq k$ the set $s_l \vdash \dlangle A \drangle  \nxt  \dlangle A \drangle \always \psi, S_l$ is true. Applying
Lemma  \ref{rev_prop_lemma} we get  that each vertex $s_i \vdash  \always \psi, U_i$ is true, hence in particular $s_1 \vdash  \always \psi, U_1$ is true. 
 
 \bigskip
 
\noindent
\textit{Inductive Step}. Here $BD(E_j)= d > 1$.

 The reasoning is essentially the same as in the basis, with the only differences:
 \begin{itemize}
 \item We reason on the structure of a subgraph rooted at $E_j = B' \cup U_1$ where $B'$ has  
 modal depth $d > 1$;
 \item The role played by Fact (a) in the basis is now played by the inductive hypothesis.
 \end{itemize}
 
\bigskip
\noindent
$\Box$


\subsubsection*{Proof of Theorem \ref{comple_th} : from right to left implication.}

What we need prove here is: \textit{if an assertion $a$ is true and $\cal P$ is a candidate proof for $a$ then $\cal P$ is indeed a proof}.

\medskip

We prove a more general property:

\noindent
\textit{Let $E$ be a set of assertions that is true  respect to a given CGM \cgm,  and let  $s \vdash_{\cgm} \varphi$ be one of its true assertions. Let $\cal P$ be a candidate proof rooted at  $E$ and
let $\lambda$ be no matter which path in it rooted at $E$. Then  $\lambda$ is successful}.  

\medskip

Let us remind that    $\lambda$ is fair by hypothesis, hence each occurrence of each assertion gets its chance of being expanded (unless $(True)$ is applied, thereby halting the construction of $\lambda$), without getting bogged in a cycle preventing any analysis.

We prove the result by induction on $\varphi$. In the induction  we consider, by an abuse of language, that if $\Psi$ is a proper sub-formula of $\Phi$ then 
$Q\;  \Psi$ is a proper sub-formula of $Q \; \Phi$, where $Q$ is the same strategic quantifier in the two formulae.\footnote{To be precise, we make the induction considering an appropriate specific notion of \textit{quasi sub-formula} of a formula, properly including sub-formulae.}

\noindent
\textit{Basis}. Here $\varphi$ is a literal, and the rule $(True)$ applies.

\noindent
\textit{Induction Step}. We must distinguish cases. 
\begin{itemize} 
 \item 

 If $\varphi$ is an $\alpha$ formula $\varphi_1 \wedge \varphi_2$, then  an application of of the $\alpha$ rule generates two expansions: $s \vdash_{\cgm} \varphi_1$  and  $s \vdash_{\cgm} \varphi_2$, one of which is in $\lambda$. It suffices to apply 
 the induction hypothesis.
\item 

 If $\varphi$ is a $\beta$ formula $\varphi_1 \vee \varphi_2$, then  an application of of the $\beta$ rule generates in $\lambda$ an expansion containing both the assertion $s \vdash_{\cgm} \varphi_1$
 and $s \vdash_{\cgm} \varphi_2$, and at least one of them is true, so
the induction hypothesis applies. Say that $s \vdash_{\cgm} \varphi_1$ is true: its proof  induces a proof for $s \vdash \varphi_1 \vee \varphi_2$\footnote{It is worthwhile recalling that, by the second item of Lemma \ref{rev_prop_lemma}, the order of application of the expansion rules in $\lambda$ does not matter, provided that $\lambda$ is fair.}
\item
When $\varphi$ is a successor  formula $Q \nxt \psi$, let us consider the case where $Q$ is an existential quantifier $\dlangle A \drangle$, the case for the universal quantifier 
being similar. The hypothesis that the assertion $s \vdash_\cgm \dlangle A \drangle \nxt \psi$ is true  implies that there is at least one collective action of the coalition $A$,
say $a$,  such that any global move completing $a$ leads to some successor of $s$ where $\psi$ is true. In other words, there exist an action $a$ such that the (meta) conjunction
${{Succes}^{A, \psi_j, s, a}}$ defined in Section \ref{next_rule} is true. Let ${{Succes}^{A, \psi_j, s, a}}$ be $s_1 \vdash \psi \; AND \;  \cdots \;  AND  \; s_n \vdash \psi$ where for
$1 \leq i \leq n$  each $s_i$ is a successor of $s$ in \cgm. The construction of the (Next)-rule, distributing conjunctions over disjunctions, assures that some $s_i \vdash \psi$
will be an element of the expansion of $s \vdash_\cgm \dlangle A \drangle \nxt \psi$ in $\lambda$. 
Hence the inductive hypothesis applied to $\psi$ assures that $\lambda$ is successful.


\item If $\varphi$ is a $\gamma$-formula  $Q \; \Phi$ we must distinguish sub-cases.

\begin{enumerate}
\item $\Phi$ has the form $\always \psi$. Then an application of the $\gamma$-rule generates in $\lambda$ an expansion containing  either $s \vdash \psi$ or   $s \vdash Q \nxt Q \always \psi$, and both these assertions are  true (remember that branching is conjunctive). In the case where $s \vdash \psi$ is present it suffices to apply the inductive hypothesis. Otherwise, an application of $(Next)$ generating a successor of the state $s$ may occur, creating a maximal cycle where $Q \; \always \psi$ is generated over and over. If the label of  the entry point of such a cycle contains 
$s \vdash Q \always \psi$ as element, the cycle does not lead to failure. 

\item $\Phi$ is an until formula $\phi_1 \until \phi_2$.
\begin{enumerate}
\item $Q$ is $\dlangle A \drangle$.\\
  Let $\lambda$ be any path in $\cal P$ rooted at a set of assertions $E$ containing the true assertion  $s \vdash \dlangle A \drangle \phi_1 \until \phi_2$. Since 
  $s \vdash \dlangle A \drangle \phi_1 \until \phi_2$ is true,   there is a strategy for the coalition $A$ such that the tree of the paths in the CGM \cgm that are compliant with that strategy
and are rooted at $s$  is such that each path $\pi$ in it has the form  $s=s_1, s_2, ...., s_k,....$ for some
$k \geq 1$,  $s_k \models \phi_2$, and $\phi_1$ is true at each $s_j$ in $[s_1,...,s_{k-1]}$; without loss of generality we can assume that $k$ is the first state where $\phi_2$ is true. Then such a path $\pi$ in the model is represented by  $\lambda$ in the sense of Lemma \ref{speriamo}. As a consequence, either for some $i$, $1 <  k$ there is a node  in $\lambda$ whose label is $E', s_i \vdash \phi_1$, for some set of assertions $E'$, or $\lambda$ contains a node whose label is $E", s_k \vdash \phi_2$, for some set of assertions $E"$.
In both cases, the inductive hypothesis (applied to $\phi_1$ or $\phi_2$) allows us to conclude that $\lambda$ is successful.
\item $Q$ is $[[A]]$.\\
Again, an application of Lemma \ref{speriamo} does the required job.

\end{enumerate}

\item 
 $\Phi$ is  $\Phi_1 \wedge \Phi_2$. In this case by construction of the $\gamma$-rule 
and Lemma \ref{rev_prop_lemma},
$Q \; \Phi$ is expanded in $\lambda$ to a set of assertions whose truth logically follows from $s \vdash_\cgm Q \; \Phi$. Let $s \vdash \psi$ be a true element of this set.   The only case in which   $\psi $ might fail to be a strict sub-formula of   $\varphi = Q \; \Phi$, to which the induction hypothesis can apply,  so to get easily the desired result,    is when  both the main operators of  $\Phi_1$ and $\Phi_2$ are in
$\{\until, \always\}$. We have three cases:
\begin{enumerate}
\item $\Phi_1$ is $\phi_1 \until \phi_2$, $\Phi_2$ is $\phi_3 \until \phi_4$, and $\psi$  has the form $Q \nxt Q \Psi$ where $\Psi$ is the conjunction  of sub-formulae in
$\phi_1 \until \phi_2$,  $\phi_3 \until \phi_4$,   for instance  $\psi$ is  $Q \nxt Q ((\phi_1 \until \phi_2) \wedge (\phi_3 \until \phi_4))$;
\item $\Phi_1$ is $\phi_1 \until \phi_2$, $\Phi_2$ is $\always \phi_3$, and $\psi$      has the form $Q \nxt Q \Psi$ where $\Psi$ is the conjunction  of sub-formulae in
$\phi_1 \until \phi_2$,  $ \always \phi_3$,   for instance $\psi$  is  is $Q \nxt Q ((\phi_1 \until \phi_2) \wedge \always \phi_3)$ (or the symmetrical case); 
\item $\Phi_1$ is  $\always \phi_1$, $\Phi_2$ is $ \always \phi_2$, and $\psi$    has the form $Q \nxt Q \Psi$,  where $\Psi$ is the conjunction  of sub-formulae in
$\always \phi_1$,  $\always \phi_2$,  for instance $\psi$  is $Q \nxt Q ((\always \phi_1) \wedge (\always \phi_2)$.

\end{enumerate}

We detail here only the first case.

\medskip

Thus, let $Q \; \Phi$ be $\dlangle A \drangle  (( \phi_1 \until \phi_2) \wedge \phi_3 \until \phi_4)$ (the case where $Q$ is universal is quite similar) and let $\psi= \dlangle A \drangle  \nxt \dlangle A \drangle  ((\phi_1 \until \phi_2) \wedge (\phi_3 \until \phi_4))$ be a true element of the expansion of $Q \; \Phi$ created by the $\gamma$-rule in $\lambda$. \\

Let $\stratA$ be any strategy for the coalition $A$ witnessing the truth of  such $ Q \; \Phi$
 in \cgm, that is, such that  any branch $\pi$ of  the tree ${\cal T}_{\cgm, \stratA}$
of the paths in the CGM \cgm that are compliant with 
$\stratA$ and are
 rooted at $s$  has the property: \\
it has the form  $s=s_1, s_2, ...., s_{k1},....,s_{k2}...$ for some
$k 1, k2 \geq 1$ where $k1 \leq k2$,  $s_{k1} \models \phi_2$,  $s_{k2} \models \phi_4$, $\phi_1$ is true at each $s_j$ in $\{s_1,...,s_{k-1}\}$ and
 $\phi_3$ is true at each $s_l$ in $\{s_1,...,s_{k2-1}\}$\\
 (or the symmetric property, swapping $k1$ and $k2$).\\

 Without any loss of generality we can suppose that $k1, k2$ are the least indexes having such a property.

\medskip

Reasoning in a way similar to Lemma \ref{speriamo} one sees that our proof path $\lambda$ must eventually contain a vertex whose label contains a true assertion  $a$ 
that is either $s_i \vdash 
\phi_1$, or $s_j \vdash 
\phi_3$, where $ 1 \leq i \leq k1-1$ and  $ 1 \leq j \leq k2-1$,  or $s_{k1} \vdash  \phi_2$ or $s_{k1}  \vdash \phi_4$. We can therefore apply the inductive hypothesis to conclude that 
$\lambda$ is successful.


 \item 
 $\Phi$ is  $\Phi_1 \vee \Phi_2$. The reasoning is similar to the case where
 $\Phi$ is  $\Phi_1 \wedge \Phi_2$.

 \end{enumerate}

 \end{itemize}
 
 \noindent
 $\Box$

\bigskip

\begin{theorem}[Termination]
\label{termination}
Given a CGM \cgm and an assertion $a = s \vdash_{\cgm} Q\; \varphi$, a candidate proof for $a$ has 
at most $2^{2^{{\mid \varphi \mid ^2}} \times m }$ 
vertices, where $m$ is the number of states in $\cgm$
\end{theorem}.

 \subsection*{Proof of Theorem \ref{termination}.}


Let $\varphi$ be an \ATLp state formula. 
Each formula in $cl(\varphi)$ has length less than 2$\mid \varphi \mid$ and is built from  symbols in $\varphi$,  so there are at most $\mid \varphi \mid^{2\mid \varphi \mid}$ such formulae  and  the size of  $cl(\varphi)$ is inferior to $2^{{\mid \varphi \mid ^2}}$.
Each formula in   $cl(\varphi)$ can be combined with $m$ states to produce an assertion, where $m=\mid \states \mid$ in $\cgm$. Hence, by Lemma \ref{closure}, the
number of distincts sets of assertions in a candidate proof is upper bounded by
$2^{2^{{\mid \varphi \mid ^2}} \times m }$.

\bigskip
\noindent
$\Box$

\section{Conclusions and related work}
\label{conclu}
In this work we have proposed $\calcul$,  a proof system for the truth of an \ATLp  formula at a given state of a model (a CGM)  as the basis for a terminating, sound and complete algorithm to perform on-the-fly model checking. Besides the already cited works on tableaux that decide satisfiability for logics of the $\ATL$ family \cite{ijcar14,atlplusjournal,davidCADE}, our work is also partially inspired by \cite{ctls}. However strategic path quantifiers, differently from path quantifiers is \CTLs, hide an alternate first order quantification and this makes formulae analysis in \ATLs particularly involved. As a consequence, the representation of  assertions in $ \calcul$ is quite different from \cite{ctls}, and this holds also for   proof vertices: in our case a vertex describes many states at the same time. The analysis is still more difficult in the case of \ATLs, and we are presently working on the extension of our approach to that highly expressive logic.

The complexity analysis underlying our proof of Theorem \ref{termination} is quite rough, and it does not take into account the fact that the branches of a candidate prof are constructed depth-first, hence that at any given state of the construction only the current branch is kept in memory.  All that is done is to provide an upper bound on the global size of a candidate proof. A more refined analysis would be needed, maybe inspired by what is done in \cite{David2015TowardsSO}, section 4.5.2. There, a fine analysis is provided  to prove that the complexity of the proposed algorithm to decide the satisfiability of an \ATLs formula is 2EXPTIME, thus optimal for the corresponding decision problem, notwithstanding the fact that a rough analysis, similar to the one proposed here for $\calcul$, would allow only to conclude that the number of vertices of a  \ATLs tableau is a 3-exponential function the size of the formula in the worst case. Therefore, we cannot conclude to optimality of our algorithm based on $\calcul$ to solve the model-checking problem for \ATLp,  that has a PSPACE complexity, and future work is needed, both to possibly improve the procedure for branch construction and to refine its complexity analysis. 

What we can already observe, however,  is that the double exponential in the upper bound on the  size of $\cal P$
comes  from the size of $\varphi$ (because $\mid cl(\varphi)$ is already an exponential function of  the size of $\varphi$),  not from the size of $\cgm$, while it is the latter the responsible of the practical problem  known as state explosion problem. 
The specific feature of our algorithm is that while constructing a path in $\cal P$ to test the truth of an assertion $s_{\cal M} \vdash \phi$, it builds only those states that
are reachable from $s$ and  necessary to  evaluate $\phi$ at $s$, and in practical cases the number of these states 
can be much smaller than the global number of states in \cgm. Even  in the  worst  performance situation,  where the tested assertion is true and the whole candidate proof $\cal P$ needs to be constructed, one can reasonably conjecture that in most of the cases $\cal P$ will have a size smaller than its theoretical upper bound, while when its root reveals false, 
the candidate proof construction can halt without building the whole candidate proof.

Clearly, besides a thorough complexity analysis,  an implementation of our method is  needed, accompanied by practical experiments allowing for
benchmarking and  studying the  performance of the algorithm on classes of assertions. It might be particularly useful to compare our method with the one 
underling the proof of the PSPACE upper bound for the model checking problem of \ATLp given in \cite{GS}. There it is shown that a formula $\phi$ holds at a state $s$ of
$\cgm$ according to the compositional semantics if and only if --in game semantic terms-- Eloise has a positional winning strategy in the evaluation game ${\cal G}(\cgm, s, \phi, N)$, where N is $\mid \cgm \mid * \mid \phi \mid$; then, the authors observe that ``it is  routine to construct an alternating Turing machine TM that simulates ${\cal G}(\cgm, s, \phi, N)$". Thus, in order to make the comparaison possible,  the game semantics evaluation algorithm  should be fully described and  receive an  implementation. It would be particularly interesting to compare the practical performance
of our method with the game semantics based method for the tractable fragments $\ATL^k$ defined in  \cite{GS}, but
these comparisons require a good amount of work and are left to future research.

\bibliographystyle{alpha}
\bibliography{ATL_plus.bib}


\end{document}